\documentclass[pre,showpacs,showkeys,preprintnumbers,amsmath,amssymb,superscriptaddress,onecolumn]{revtex4}
\usepackage{graphicx}
\usepackage{amsfonts}
\usepackage{url}
\usepackage{epstopdf}
\usepackage{color}
\usepackage{bm}

\def\L{\mathcal L}
\def\N{\mathcal N}
\def\var{{\rm var}}

\def\s{\bm{s}}
\def\x{\bm{x}}
\def\X{\bm{X}}

\def\pa{\partial\Omega}
\def\C{{\mathbb C}}
\def\E{{\mathbb E}}

\def\P{{\mathbb P}}
\def\R{{\mathbb R}}

\def\L{{\mathcal L}}

\def\M{{\mathcal M}}
\def\I{{\mathcal I}}
\def\K{{\mathcal K}}

\def\erf{\mathrm{erf}}
\def\erfc{\mathrm{erfc}}

\def\ctanh{\mathrm{ctanh}}
\def\sech{\mathrm{sech}}

\begin{document}

\title{Joint distribution of multiple boundary local times 
\\ and related first-passage time problems with multiple targets}

\author{Denis~S.~Grebenkov}
	\email{denis.grebenkov@polytechnique.edu}
\affiliation{Laboratoire de Physique de la Mati\`{e}re Condens\'{e}e (UMR 7643), \\ 
CNRS -- Ecole Polytechnique, IP Paris, 91128 Palaiseau, France}
\affiliation{Institute for Physics and Astronomy, University of Potsdam, 14476 Potsdam-Golm,
Germany}

\begin{abstract}
We investigate the statistics of encounters of a diffusing particle
with different subsets of the boundary of a confining domain.  The
encounters with each subset are characterized by the boundary local
time on that subset.  We extend a recently proposed approach to
express the joint probability density of the particle position and of
its multiple boundary local times via a multi-dimensional Laplace
transform of the conventional propagator satisfying the diffusion
equation with mixed Robin boundary conditions.  In the particular
cases of an interval, a circular annulus and a spherical shell, this
representation can be explicitly inverted to access the statistics of
two boundary local times.  We provide the exact solutions and their
probabilistic interpretation for the case of an interval and sketch
their derivation for two other cases.  We also obtain the
distributions of various associated first-passage times and discuss
their applications.
\end{abstract}

\pacs{02.50.-r, 05.40.-a, 02.70.Rr, 05.10.Gg}

%02.50.-r       (Probability theory, stochastic processes, and statistics)
%05.40.-a 	Fluctuation phenomena, random processes, noise, and Brownian motion
%02.70.Rr       (General statistical methods)
%05.10.Gg 	Stochastic analysis methods (Fokker-Planck, Langevin, etc.) 

%02.50.Ey 	Stochastic processes  (Probability theory, stochastic processes, and statistics)

\keywords{Diffusion-influenced reactions, Boundary local time, Statistics of encounters, 
Surface reaction, Robin boundary condition, Heterogeneous catalysis}

%\keywords{Diffusion-influenced reactions, Boundary local time, Statistics of encounters, 
%Surface reaction, Robin boundary condition, Heterogeneous catalysis}

%\submitto{\JSTAT}

\date{\today}

\maketitle

\section{Introduction}

Diffusion-controlled reactions and related stochastic processes in an
Euclidean domain $\Omega\subset \R^d$ are typically described by the
propagator (also known as the heat kernel or the Green's function),
$G_q(\x,t|\x_0)$, that is the probability density of the event that a
particle started from $\x_0\in \Omega$ at time $0$ has arrived in a
vicinity of a point $\x \in\Omega$ at time $t$ without being killed
(or reacted) on the boundary $\pa$ of the domain
\cite{Gardiner,Redner,Schuss}.
For ordinary diffusion, this propagator satisfies the diffusion
equation (for any starting point $\x_0 \in \Omega$),
\begin{equation}  \label{eq:diff}
\partial_t G_q(\x,t|\x_0) = D \Delta G_q(\x,t|\x_0)  \qquad (\x\in \Omega),
\end{equation}
subject to the initial condition $G_q(\x,t=0|\x_0) =
\delta(\x-\x_0)$ and the Robin boundary condition on $\pa$:
\begin{equation}  \label{eq:Robin_1}
\partial_n G_q(\x,t|\x_0) + q\, G_q(\x,t|\x_0) = 0 \qquad (\x\in\pa), 
\end{equation}
where $D$ is the diffusion coefficient, $\Delta$ is the Laplace
operator (acting on $\x$), $\delta(\x-\x_0)$ is the Dirac
distribution, and $\partial_n$ is the normal derivative on the
boundary $\pa$, oriented outward the domain $\Omega$.  The parameter
$q$ characterizes the reactivity of the boundary and ranges from $q =
0$ (an inert reflecting boundary with Neumann condition) to $+\infty$
(a perfectly reactive boundary with Dirichlet condition).  The
intermediate case of $0 < q < +\infty$ corresponds to partial
reactivity of the boundary which can represent overpassing a reaction
activation barrier
\cite{Collins49,Sano79,Sano81,Hanggi90,Zhou91,Reguera06,Grebenkov17},
the coarse-graining effect of microscopic spatial heterogeneities of
reactive patches
\cite{Berg77,Zwanzig90,Zwanzig91,Berezhkovskii04,Berezhkovskii06,Muratov08,Skvortsov14,Skvortsov15,Dagdug16,Lindsay17,Bernoff18b,Skvortsov19},
stochastic activity of the target (open/closed channels, switching
between conformational states of a macromolecule)
\cite{Benichou00,Reingruber09,Lawley15,Bressloff17}, and other
microscopic mechanisms
\cite{Brownstein79,Sapoval94,Filoche99,Sapoval02,Grebenkov05,Grebenkov06a,Grebenkov07a,Grebenkov10a}
(see an overview in \cite{Grebenkov19b}).  The propagator determines
most commonly employed characteristics of diffusion-reaction processes
such as the survival probability, the reaction time distribution, and
the reaction rate, that found numerous applications in physics,
chemistry and biology
\cite{Rice,Lauffenburger,Metzler,Oshanin,Bouchaud90,Grebenkov07,Benichou11,Bressloff13,Bray13,Benichou14}.
Importantly, the propagator and all related quantities depend on $q$
{\it implicitly} (as a parameter of the boundary condition) that makes
the study of this dependence and its eventual optimization
challenging.

To overcome this limitation, we proposed an alternative description of
partial reactivity in terms of the boundary local time that quantifies
the encounters of a diffusing particle with the boundary of a
confining domain \cite{Grebenkov20}.  The boundary local time $\ell_t$
naturally appears in the stochastic differential equation for
reflected Brownian motion $\X_t$ \cite{Levy,Ito,Freidlin} and can be
expressed in terms of the residence time of $\X_t$ in a thin boundary
layer $\pa_a$
\begin{equation}
\ell_t = \lim\limits_{a\to 0} \frac{D}{a} \underbrace{\int\limits_0^t dt' \, \Theta(a - |\pa - \X_{t'}|)}_{\textrm{residence time in}~\pa_a} ,
\end{equation}
where $\Theta(z)$ is the Heaviside step function, which represents the
indicator function of a thin layer of width $a$ near $\pa$: $\pa_a =
\{ \x\in\Omega ~:~ |\x - \pa| < a\}$.  Note that the prefactor $D/a$
renders $\ell_t$ to be in units of length.  We also stress that
the boundary local time should not be confused with the point local
time, which was thoroughly studied in the past (see
\cite{Borodin,Takacs95,Randon18} and references therein).
For reflected Brownian motion on the half-line, the distribution of
the boundary local time has been studied long ago \cite{Levy,Borodin}.
In a recent paper, we proposed a general spectral approach to obtain
the distribution of the boundary local time for Euclidean domains with
smooth boundary by using the Dirichlet-to-Neumann operator
\cite{Grebenkov19c}.  This approach was further extended in
\cite{Grebenkov20} to get the joint probability density
$P(\x,\ell,t|\x_0)$ of the position $\X_t$ of the particle diffusing
in a domain $\Omega$ with {\it reflecting} boundary, and of its
boundary local time $\ell_t$ at time $t$, given that it has started
from a point $\x_0$ at time $0$.  This so-called full propagator was
shown to be related to the conventional propagator $G_q(\x,t|\x_0)$
via the Laplace transform:
\begin{equation}  \label{eq:GP_one}
G_q(\x,t|\x_0) = \int\limits_0^\infty d\ell \, e^{-q\ell} \, P(\x,\ell,t|\x_0).
\end{equation}
Here, the surface reactivity parameter $q$ appears explicitly in the
prefactor $e^{-q\ell}$ which comes from the assumption of constant
boundary reactivity.  Other reaction mechanisms with
encounter-dependent reactivity were introduced and studied in
\cite{Grebenkov20} (see also Sec. \ref{sec:conclusion} below).
We emphasize that the propagators $G_q(\x,t|\x_0)$ (with $q >
0$) and $G_0(\x,t|\x_0)$ (or its extension $P(\x,\ell,t|\x_0)$)
correspond to the distinct cases of reactive and reflecting (inert)
boundaries, respectively.  As the particle may react and thus
disappear in the former case, the associated diffusive processes are
usually distinguished in the literature.  However, as argued in
\cite{Grebenkov20} (see also
\cite{Grebenkov06a,Grebenkov07a,Grebenkov19b} and references therein),
the diffusive process in the presence of a reactive boundary is just
reflected Brownian motion in a domain with reflecting boundary, which
is stopped at an appropriate random time.  This property is reflected
by Eq. (\ref{eq:GP_one}), in which the full propagator
$P(\x,\ell,t|\x_0)$ characterizes reflected Brownian motion and the
prefactor $e^{-q\ell}$ incorporates the stopping condition (see
Sec. \ref{sec:general} below for details and extensions).

In many chemical and biological applications, the reactive boundary is
not homogeneous, while a reactive patch or a target is not unique.
For instance, many traps can compete for capturing the diffusing
particle, and one is interested in knowing the capture time for a
given trap in the presence of its competitors \cite{Grebenkov20a}.
Even a single trap can be surrounded by inert obstacles or by a
reflecting boundary.  When considering an escape problem, the escape
region is usually a subset of the reflecting boundary.  In all these
situations, setting the homogeneous Robin boundary condition
(\ref{eq:Robin_1}) on the whole boundary is not appropriate, as one
has to distinguish surface mechanisms on different regions of the
boundary.  For this purpose, a spectral approach with a
space-dependent reactivity was developed \cite{Grebenkov19}.

In this paper, we propose a complementary approach and bring some
probabilistic insights onto this problem when the reactivity is
piecewise constant.  In this case, one can consider different
reactivity regions by partitioning the boundary $\pa$ into $m$
non-overlapping subsets $\Gamma_i$:
\begin{equation}
\pa = \bigcup_{i=1}^m \overline{\Gamma}_i,  \qquad \Gamma_i \cap \Gamma_j = \emptyset.
\end{equation}
In order to characterize the encounters with different parts
$\Gamma_i$ of the boundary, we introduce the associated boundary local
times $\ell_t^i$:
\begin{equation}
\ell_t^i = \lim\limits_{a\to 0} \frac{D}{a} \int\limits_0^t dt' \, \Theta(a - |\Gamma_i - \X_{t'}|)  \qquad (i=1,\ldots,m).
\end{equation}
If the joint distribution of the boundary local times $\ell_t^i$ was
known, one could investigate various encounter properties such as
``How many times the particle has arrived on a given partially
reactive trap before being absorbed by its competitors?'', ``What is
the first moment when the particle has visited each trap a given
number of times?'', etc.  In other words, the joint distribution of
the boundary local times will provide conceptually new insights onto
diffusion-controlled reactions, far beyond the conventional
first-passage times.  To our knowledge, such joint distributions were
not studied earlier.

In this paper, we aim at obtaining the joint distribution by extending
the probabilistic arguments from \cite{Grebenkov20}.  In
Sec. \ref{sec:general}, we generalize Eq. (\ref{eq:GP_one}) to a
multi-dimensional Laplace transform and discuss some of its
properties.  However, the numerical inversion of the multi-dimensional
Laplace transform is challenging.  For this reason, we restrict our
attention to three basic domains (an interval, a circular annulus and
a spherical shell) for which the inversion can be performed explicitly
(Sec. \ref{sec:interval}).  In fact, we derive an exact formula for
the joint probability density for the case of an interval, and discuss
its straightforward extension for two other domains.  We illustrate
the properties of the two boundary local times and their correlations.
Section \ref{sec:FPT} is devoted to various first-passage time
problems.  We first recall the basic first-passage times to perfectly
and partially reactive boundary and then derive the probability
density of the first time when two boundary local times exceed
prescribed thresholds.  In other words, we fully characterize the
first moment when both subsets of the boundary have been visited a
prescribed number of times.  In Sec. \ref{sec:conclusion}, we discuss
some further extensions and consequences of the obtained results for
diffusion-controlled reactions.

\section{General solution}
\label{sec:general}

The joint distribution can be derived by extending the probabilistic
arguments from \cite{Grebenkov20}.  For this purpose, let us introduce
the propagator $G_{q_1,\ldots,q_m}(\x,t|\x_0)$ satisfying the
diffusion equation (\ref{eq:diff}) with mixed Robin boundary
conditions:
\begin{equation}  \label{eq:Robin}
\partial_n G_{q_1,\ldots,q_m}(\x,t|\x_0) + q_i\, G_{q_1,\ldots,q_m}(\x,t|\x_0) = 0 \qquad (\x\in\Gamma_i),
\end{equation}
with nonnegative parameters $q_1,\ldots,q_m$ characterizing each
reactive part $\Gamma_i$ of the boundary.  In other words, we extend
the constant reactivity parameter $q$ from Eq. (\ref{eq:Robin_1}) by a
piecewise constant function taking the values $q_1,\ldots,q_m$ on
different subsets $\Gamma_i$ of the boundary.  As discussed in
\cite{Grebenkov06,Grebenkov06b,Grebenkov07a,Grebenkov19b}, the Robin
boundary condition describes partial reactivity of the boundary: the
diffusing particle hitting the boundary can either react, or be
reflected.  To define properly the reaction probability $\Pi$ at each
encounter, one can introduce a thin layer of width $a$ near the
reactive part $\Gamma_i$, for which $\Pi_i = a q_i/(1 + a q_i)$ (and
if the particle is not reacted, it is reflected at distance $a$ from
the boundary).  For a finite $q_i$ and small $a$, one has $\Pi_i
\approx a q_i$.  In the limit $a\to 0$, the probability of the reaction
event goes to $0$ but the number of returns to the boundary goes to
infinity, yielding a nontrivial limit.  If all attempts to react are
independent, the probability of not reacting on the boundary up to
time $t$ is
\begin{equation}
{\mathcal P}_t = \E_{\x_0}\biggl\{ \prod\limits_{i=1}^m (1 - \Pi_i)^{\N_{t,a}^i} \biggr\} ,
\end{equation}
where $\N_{t,a}^i$ is the number of encounters with a thin layer near
$\Gamma_i$ up to time $t$, and $\E_{\x_0}$ denotes the expectation
with respect to the probability measure associated with reflected
Brownian motion in $\Omega$, started from $\x_0$.  In the limit $a\to
0$, this number is related to the boundary local time: $\N_{t,a}^i
\approx \ell_t^i/a$ \cite{Levy,Ito,Freidlin} so that
\begin{equation} 
{\mathcal P}_t \approx  \E_{\x_0}\biggl\{ \exp\biggl(- \sum\limits_{i=1}^m \Pi_i \N_{t,a}^i \biggr) \biggr\} \xrightarrow[a\to 0]{}
\E_{\x_0}\biggl\{ \exp\biggl(- \sum\limits_{i=1}^m q_i \ell_{t}^i \biggr) \biggr\} .
\end{equation}
Adding the constraint for the arrival position of the particle to be
in a vicinity of $\x$, one gets the probabilistic meaning of the
conventional propagator, i.e., the probability density of finding the
survived particle in a vicinity of $\x$:
\begin{equation}
G_{q_1,\ldots,q_m}(\x,t|\x_0) = \E_{\x_0}\biggl\{ \exp\biggl(- \sum\limits_{i=1}^m q_i \ell_{t}^i \biggr)  \delta(\X_t - \x) \biggr\} .
\end{equation}
If $P(\x,\ell_1,\ldots,\ell_m,t|\x_0)$ is the joint probability
density of the position $\X_t$ and of all boundary local times
$\ell_t^i$, the above expectation reads
\begin{equation}  \label{eq:GP_general}
%\fl \qquad
G_{q_1,\ldots,q_m}(\x,t|\x_0) = \int\limits_0^\infty d\ell_1 \, e^{-q_1\ell_1}\ldots \int\limits_0^\infty d\ell_m \, e^{-q_m\ell_m} \, 
P(\x,\ell_1,\ldots,\ell_m,t|\x_0).
\end{equation}
This is the extension of Eq. (\ref{eq:GP_one}) derived in
\cite{Grebenkov20}.  Formally, the joint probability density of the
boundary local times $\ell_t^1, \ldots, \ell_t^m$ and of the position
$\X_t$ can be obtained from the propagator
$G_{q_1,\ldots,q_m}(\x,t|\x_0)$ by performing the $m$-fold Laplace
transform inversion.

The (marginal) joint probability density of the boundary local times
$\ell_t^1, \ldots, \ell_t^m$ is simply
\begin{equation}
P(\circ, \ell_1,\ldots,\ell_m,t|\x_0) = \int\limits_{\Omega} d\x \, P(\x,\ell_1,\ldots,\ell_m,t|\x_0)
\end{equation}
(we use the notation $\circ$ for marginalized variables).  Integrating
Eq. (\ref{eq:GP_general}) over $\x\in\Omega$, one gets
\begin{equation}  \label{eq:Srho_general}
S_{q_1,\ldots,q_m}(t|\x_0) = \int\limits_0^\infty d\ell_1 \, e^{-q_1\ell_1}\ldots \int\limits_0^\infty d\ell_m \, e^{-q_m\ell_m} \, 
P(\circ,\ell_1,\ldots,\ell_m,t|\x_0),
\end{equation}
where 
\begin{equation}
S_{q_1,\ldots,q_m}(t|\x_0) = \int\limits_{\Omega} d\x \, G_{q_1,\ldots,q_m}(\x,t|\x_0)
\end{equation}
is the survival probability up to time $t$ in the presence of reactive
traps, which obeys the backward diffusion equation:
\begin{subequations}
\begin{align}
\partial_t S_{q_1,\ldots,q_m}(t|\x_0) & = D \Delta S_{q_1,\ldots,q_m}(t|\x_0)  \qquad (\x_0 \in \Omega), \\  \label{eq:S_Robin}
\partial_n S_{q_1,\ldots,q_m}(t|\x_0) + q_i\, S_{q_1,\ldots,q_m}(t|\x_0) & = 0 \qquad (\x_0 \in\Gamma_i), 
\end{align}
\end{subequations}
subject to the initial (terminal) condition
$S_{q_1,\ldots,q_m}(t=0|\x_0) = 1$.  Note also that the Laplace
transform (\ref{eq:Srho_general}) allows one to determine joint
positive-order integer moments of the boundary local times:
\begin{equation}  \label{eq:moments}
%\fl
\E_{\x_0} \bigl\{ [\ell_t^1]^{k_1} \ldots  [\ell_t^m]^{k_m} \bigr\} = (-1)^{k_1+\ldots+k_m} \lim\limits_{q_1,\ldots,q_m\to 0} 
\frac{\partial^{k_1+\ldots+k_m}}{\partial q_1^{k_1} \ldots \partial q_m^{k_m}} S_{q_1,\ldots,q_m}(t|\x_0) 
\end{equation}
for any integer $k_1,\ldots,k_m \geq 0$ (in turn, using
Eq. (\ref{eq:GP_general}), one gets the moments under the additional
constraint of being in $\x$ at time $t$).

As reflected Brownian motion is a Markovian process, the conventional
propagator $G_{q_1,\ldots,q_m}(\x,t|\x_0)$ gives access to the joint
probability density of $k$ positions $\x^1, \x^2, \ldots, \x^k$ at
successive times $0 < t_1 < t_2 < \ldots < t_k$ as
\begin{equation}
%\fl \quad
G_{q_1,\ldots,q_m}(\x^1,t_1|\x_0) \, G_{q_1,\ldots,q_m}(\x^2,t_2-t_1|\x^1) \ldots G_{q_1,\ldots,q_m}(\x^k,t_k-t_{k-1}|\x^{k-1}) .
\end{equation}
The same property holds for the full propagator, which determines the
successive positions $\x^j$ and values of all boundary local times
$\ell^j_i$ ($j=1,\ldots,k$, $i = 1,\ldots,m$) at times
$t_1,t_2,\ldots,t_k$:
\begin{align} \nonumber  
& P(\x^1,\ell_1^1,\ldots,\ell_m^1,t_1|\x_0) \, P(\x^2,\ell_1^2-\ell_1^1,\ldots,\ell_m^2-\ell_m^1,t_2-t_1|\x^1) \\
& \qquad \ldots P(\x^k,\ell_1^k-\ell_1^{k-1},\ldots,\ell_m^k-\ell_m^{k-1},t_k-t_{k-1}|\x^{k-1}) .
\end{align}

Even though Eqs. (\ref{eq:GP_general}, \ref{eq:Srho_general}) give
access to the joint probability densities
$P(\x,\ell_1,\ldots,\ell_m,t|\x_0)$ and
$P(\circ,\ell_1,\ldots,\ell_m,t|\x_0)$, these expressions are in
general rather formal because the propagator
$G_{q_1,\ldots,q_m}(\x,t|\x_0)$ and the survival probability
$S_{q_1,\ldots,q_m}(t|\x_0)$ are rarely known analytically, whereas
the numerical inversion of the (multi-dimensional) Laplace transform
can be unstable \cite{Epstein08}.  For this reason, obtaining these
joint probability densities in a more constructive way (such as, e.g.,
the spectral approach in \cite{Grebenkov20}) remains an open problem.

Lacking yet a general constructive approach, we further focus on joint
probability densities for three basic domains: an interval, a circular
annulus between two concentric circles, and a spherical shell between
two concentric spheres.  The boundary of these domains naturally
splits into two disjoint parts $\Gamma_1$ and $\Gamma_2$, so that we
are limited to $m = 2$.  In two and three dimensions, the rotational
symmetry of these domains reduces the computation to a one-dimensional
setting for which the joint probability densities can be derived
analytically.  This derivation relies on the explicit form of the
propagator in the Laplace domain (with respect to time $t$, denoted by
tilde throughout the paper):
\begin{equation}
\tilde{G}_{q_1,q_2}(\x,p|\x_0) = \int\limits_0^\infty dt \, e^{-pt} \, G_{q_1,q_2}(\x,t|\x_0),
\end{equation}
which obeys the modified Helmholtz equation
\begin{equation}  \label{eq:Helm}
(p - D\Delta) \tilde{G}_{q_1,q_2}(\x,p|\x_0) = \delta(\x-\x_0)  \qquad (\x\in \Omega),
\end{equation}
with Robin boundary conditions
\begin{equation}  \label{eq:Robin2}
\left. \biggl(\partial_n \tilde{G}_{q_1,q_2}(\x,p|\x_0) + q_i \tilde{G}_{q_1,q_2}(\x,p|\x_0)\biggr)\right|_{\x\in \Gamma_i} = 0  \qquad (i=1,2).
\end{equation}
Note that the Laplace-transformed propagator also allows one to
describe diffusion-influenced reactions for mortal particles
\cite{Yuste13,Meerson15,Grebenkov17d}.  In the next section, we
present the detailed derivation for an interval, while its extension
to an annulus and a spherical shell will be sketched in
Sec. \ref{sec:extensions}.

\section{Exact explicit solution}
\label{sec:interval}

For an interval $(0,b)$, the boundary consists of two endpoints,
$\Gamma_1 = \{0\}$ and $\Gamma_2 = \{b\}$, and the Laplace-transformed
propagator satisfying Eqs. (\ref{eq:Helm}, \ref{eq:Robin2}), is known
explicitly \cite{Thambynayagam} (see \cite{Grebenkov19g} for details):
\begin{equation}   \label{eq:tildeG}
D\tilde{G}_{q_1,q_2}(x,p|x_0) = \frac{1}{\alpha V(\alpha)} \times 
\left\{ \begin{array}{l l} v^b(x_0) v^0(x)  ,& \qquad 0 \leq x \leq x_0 , \\
v^0(x_0) v^b(x)  ,& \qquad x_0 \leq x \leq b , \end{array} \right.
\end{equation}
where
\begin{align*}
v^0(x) & = q_1 \sinh(\alpha x) + \alpha \cosh(\alpha x) , \\
v^b(x) & = q_2 \sinh(\alpha (b-x)) + \alpha \cosh(\alpha (b-x)) , \\
V & = (\alpha^2 + q_1 q_2) \sinh(\alpha b) + \alpha(q_1 + q_2) \cosh(\alpha b),
\end{align*}
with $\alpha = \sqrt{p/D}$.  We aim at evaluating explicitly the
inverse double Laplace transform with respect to $q_1$ and $q_2$
(denoted as $\L_2^{-1}$) to get the full propagator:
\begin{equation}  \label{eq:PG_double}
\tilde{P}(x,\ell_1,\ell_2,p | x_0) = \L_2^{-1} \bigl\{ \tilde{G}_{q_1,q_2}(x,p|x_0) \bigr\} ,
\end{equation}
i.e., the Laplace transform (with respect to time $t$) of the joint
probability density of the position $x$ and two boundary local times
$\ell_1$ and $\ell_2$ at endpoints $\Gamma_1$ and $\Gamma_2$,
respectively.  Even though an extra Laplace inversion will be needed
to get $P(x,\ell_1,\ell_2,t|x_0)$ in time domain, this is much simpler
than the original double Laplace transform.  Moreover, it is common to
operate with diffusion characteristics in the Laplace domain, in
particular, when studying first-passage times (see below).

\subsection{Moments of the boundary local times}

Before deriving the full propagator, we start by looking at the
positive moments of two boundary local times:
\begin{equation}
M_{k_1,k_2}(t) = \E_{x_0} \bigl\{ [\ell_t^1]^{k_1} \, [\ell_t^2]^{k_2} \bigr\} .
\end{equation}
According to Eq. (\ref{eq:moments}), the Laplace transform of these
moments can be obtained by integrating the propagator
$\tilde{G}_{q_1,q_2}(x,p|x_0)$ over $x$ and differentiating with
respect to $q_1$ and $q_2$:
\begin{equation}  \label{eq:moments2}
\tilde{M}_{k_1,k_2}(p) 
= (-1)^{k_1+k_2} \lim\limits_{q_1,q_2\to 0} \frac{\partial^{k_1+k_2}}{\partial q_1^{k_1} \partial q_2^{k_2}} \tilde{S}_{q_1,q_2}(p|x_0) ,
\end{equation}
where
\begin{align} \label{eq:tildeS}
\tilde{S}_{q_1,q_2}(x,p|x_0) & = \int\limits_0^b dx \, \tilde{G}_{q_1,q_2}(x,p|x_0) = \frac{1}{D \alpha^2 V(\alpha)}   \\  \nonumber
& \times \biggl\{q_1q_2 \bigl(\sinh(\alpha b) - \sinh(\alpha(b-x_0)) - \sinh(\alpha x_0)\bigr) 
 + \alpha^2 \sinh(\alpha b)  \\      \nonumber 
& + \alpha q_1 \bigl(\cosh(\alpha b) - \cosh(\alpha(b-x_0))\bigr) 
+ \alpha q_2 \bigl(\cosh(\alpha b) - \cosh(\alpha x_0)\bigr) \biggr\} \,.
\end{align}
For instance, one gets the Laplace transform of the mean values,
\begin{equation}
\tilde{M}_{1,0}(p) = \frac{\cosh(\alpha(b-x_0))}{D\alpha^3 \sinh(\alpha b)} \,, \qquad
\tilde{M}_{0,1}(p) = \frac{\cosh(\alpha x_0)}{D\alpha^3 \sinh(\alpha b)}  \,,
\end{equation}
second moments,
\begin{equation}
%\fl \qquad
\tilde{M}_{2,0}(p) = \frac{2\cosh(\alpha b)\cosh(\alpha(b-x_0))}{D\alpha^4 \sinh^2(\alpha b)} \,, \qquad
\tilde{M}_{0,2}(p) = \frac{2\cosh(\alpha b)\cosh(\alpha x_0)}{D\alpha^4 \sinh^2(\alpha b)}  \,,
\end{equation}
and the cross-moment
\begin{equation}
\tilde{M}_{1,1}(p) = \frac{\sinh(\alpha x_0) + \sinh(\alpha(b-x_0))}{D\alpha^4 \sinh(\alpha b)(\cosh(\alpha b)-1)} \,.
\end{equation}
Even so the inverse Laplace transform of these moments can be computed
exactly by the residue theorem, we just provide the asymptotic
behavior of the mean values: 

$\bullet$ at short times,
\begin{subequations}
\begin{align}
\E_{x_0} \bigl\{ \ell_t^1 \bigr\} & \simeq \left\{ \begin{array}{l l} 
	\displaystyle  \frac{4(Dt)^{3/2} e^{-x_0^2/(4Dt)}}{\sqrt{\pi} x_0^2} & \quad (x_0 > 0) , \\
	\displaystyle  \frac{2\sqrt{Dt}}{\sqrt{\pi}}  & \quad (x_0 = 0), \end{array} \right.  \\
\E_{x_0} \bigl\{ \ell_t^2 \bigr\} & \simeq \left\{ \begin{array}{l l} 
	\displaystyle  \frac{4(Dt)^{3/2} e^{-(b-x_0)^2/(4Dt)}}{\sqrt{\pi} (b-x_0)^2} & \quad (x_0 < b), \\
	\displaystyle  \frac{2\sqrt{Dt}}{\sqrt{\pi}}  & \quad (x_0 = b); \end{array} \right.  
\end{align}
\end{subequations}

$\bullet$ at long times
\begin{equation}  \label{eq:mean_tinf}
\E_{x_0} \bigl\{ \ell_t^1 \bigr\} \simeq \frac{Dt}{b} + \frac{2b^2 - 6bx_0 + 3x_0^2}{6b} \,, \qquad
\E_{x_0} \bigl\{ \ell_t^2 \bigr\} \simeq \frac{Dt}{b} + \frac{3x_0^2 - b^2}{6b} \,.
\end{equation}
Performing the same analysis for the second moment, we get the
long-time behavior of the variance, which does not depend on $x_0$ and
$b$ in the leading order:
\begin{equation} \label{eq:var_ell}
\var \bigl\{ \ell_t^1 \bigr\} \simeq  \var \bigl\{ \ell_t^2 \bigr\} \simeq 2Dt + O(1)  \qquad (t\to\infty).
\end{equation}
Finally, we get
\begin{equation}
M_{1,1} \simeq - Dt/3 + O(1)  \qquad (t\to\infty),
\end{equation}
so that the correlation between two boundary local times approaches
$-1/6$ at long times.  As expected, the correlation is negative: when
$\ell_t^1$ is larger than its mean, the particle spent more time on
the endpoint $\Gamma_1$, and thus $\ell_t^2$ is expected to be smaller
than its mean.

\subsection{Derivation of the full propagator}

Let us focus on the case $0 \leq x\leq x_0$ and write explicitly
\begin{equation}
D \tilde{G}_{q_1,q_2}(x,p|x_0) = Q(q_1,q_2) \frac{\sinh(\alpha (b-x_0))\sinh(\alpha x)}{\alpha \sinh(\alpha b)} \,,
\end{equation}
where
%
%\begin{widetext}
\begin{equation}
Q(q_1,q_2) = \frac{[q_2 + \alpha \, \ctanh(\alpha (b-x_0))][q_1 + \alpha \, \ctanh(\alpha x)]}
{q_1 q_2 + \alpha(q_1 + q_2) \ctanh(\alpha b) + \alpha^2}  \,.
\end{equation}
%\end{widetext}
To invert this double Laplace transform, we use the following
properties \cite{Debnath16}:
\begin{subequations}
\begin{align}  \label{eq:L2_1}
\L_2\{ e^{-k_1\ell_1 - k_2\ell_2} f(\ell_1,\ell_2) \} & = \L_2\{ f\}(q_1+k_1,q_2+k_2) \,,\\  \label{eq:L2_2}
\L_2\{ I_0(a\sqrt{\ell_1 \ell_2})\} & = \frac{1}{q_1q_2 -a^2/4} \,, \\  \label{eq:L2_3}
\L_2\{ \partial_{\ell_1} f(\ell_1,\ell_2)\} & = q_1 \L_2\{ f\} - \L\{ f(0,\ell_2)\} \,,
\end{align}
\end{subequations}
where $\L$ and $\L_2$ denote the single and double Laplace transforms
of some function $f$, and $I_\nu(z)$ is the modified Bessel function
of the first kind.

Using the first property, one can make the change $q_1 \to \bar{q}_1 =
q_1 + C$ and $q_2 \to \bar{q}_2 = q_2 + C$ with 
\begin{equation}  \label{eq:C}
C = \alpha \, \ctanh(\alpha b) ,
\end{equation}
so that
\begin{equation}
Q(q_1,q_2) = \bar{Q}(\bar{q}_1,\bar{q}_2) = \frac{(\bar{q}_2 +A)(\bar{q}_1 +B)}{\bar{q}_1 \bar{q}_2 - E^2/4}  \,,
\end{equation}
where 
\begin{subequations}
\begin{align}  \label{eq:E}
E & = 2 \sqrt{C^2 - \alpha^2} = 2\alpha \sqrt{\ctanh^2(\alpha b) - 1} = \frac{2 \alpha}{\sinh(\alpha b)} > 0 , \\
A & = \alpha \, \ctanh(\alpha (b-x_0)) - C , \\
B & = \alpha \, \ctanh(\alpha x) - C  .
\end{align}
\end{subequations}
We represent the above function as
\begin{equation} \label{eq:Fbar}
%\fl \qquad 
\bar{Q}(\bar{q}_1,\bar{q}_2) = 1 + A \frac{\bar{q}_1}{\bar{q}_1 \bar{q}_2 - E^2/4} + B \frac{\bar{q}_2}{\bar{q}_1 \bar{q}_2 - E^2/4} 
+ (E^2/4 + AB) \frac{1}{\bar{q}_1 \bar{q}_2 - E^2/4} \,.
\end{equation}
The inverse double Laplace transform of the first term yields
$\delta(\ell_1) \delta(\ell_2)$, whereas Eq. (\ref{eq:L2_3}) allows
one to compute it for the last term.  Using the properties
(\ref{eq:L2_2}, \ref{eq:L2_3}), we can also write
\begin{subequations}
\begin{align}
\L_2\biggl\{ \partial_{\ell_1} I_0(a\sqrt{\ell_1\ell_2}) \biggr\} & = 
\L_2\biggl\{ \frac{a \sqrt{\ell_2} I_1(a\sqrt{\ell_1\ell_2})}{2\sqrt{\ell_1}} \biggr\} 
= \frac{q_1}{q_1q_2 - a^2/4} - \frac{1}{q_2} \,,\\
\L_2\biggl\{ \partial_{\ell_2} I_0(a\sqrt{\ell_1\ell_2}) \biggr\} & = 
\L_2\biggl\{ \frac{a \sqrt{\ell_1} I_1(a\sqrt{\ell_1\ell_2})}{2\sqrt{\ell_2}} \biggr\} 
= \frac{q_2}{q_1q_2 - a^2/4} - \frac{1}{q_1} \,.
\end{align}
\end{subequations}
These relations allow us to invert the second and third terms in
Eq. (\ref{eq:Fbar}).  Combining these results, we get
\begin{align*}
& \L_2^{-1} \{\bar{Q}\} = \delta(\ell_1) \delta(\ell_2) + A \biggl(\frac{E \sqrt{\ell_2} I_1(E \sqrt{\ell_1 \ell_2})}{2\sqrt{\ell_1}} 
+ \delta(\ell_1)\biggr)  \\
& + B \biggl(\frac{E \sqrt{\ell_1} I_1(E \sqrt{\ell_1\ell_2})}{2\sqrt{\ell_2}} 
+ \delta(\ell_2)\biggr) + (E^2/4 + AB) I_0\bigl(E \sqrt{\ell_1 \ell_2}\bigr) ,
\end{align*}
from which the full propagator reads (for $0 \leq x\leq x_0 \leq b$):
\begin{equation}
D \tilde{P}(x,\ell_1,\ell_2,p|x_0) = \frac{\sinh(\alpha (b-x_0))\sinh(\alpha x)}{\alpha \sinh(\alpha b)} e^{-C(\ell_1 + \ell_2)} \L_2^{-1}\{ \bar{Q}\} \,.
\end{equation}
After simplifications, this relation becomes
\begin{align}  \nonumber
& \tilde{P}(x,\ell_1,\ell_2,p|x_0) = \underbrace{\frac{\sinh(\alpha (b-x_0))\sinh(\alpha x)}{D\alpha \sinh(\alpha b)}
}_{=\tilde{G}_{\infty,\infty}(x,p|x_0)} \delta(\ell_1) \delta(\ell_2) \\  \nonumber
& + \underbrace{\frac{\sinh(\alpha x)}{\sinh(\alpha b)}}_{=\tilde{j}_{\infty,\infty}(b,p|x)} \delta(\ell_1) \frac{e^{-C\ell_2}}{D}
 \underbrace{\frac{\sinh(\alpha x_0)}{\sinh(\alpha b)}}_{=\tilde{j}_{\infty,\infty}(b,p|x_0)}  
 + \underbrace{\frac{\sinh(\alpha (b-x))}{\sinh(\alpha b)}}_{=\tilde{j}_{\infty,\infty}(0,p|x)} 
\delta(\ell_2) \frac{e^{-C\ell_1}}{D}
  \underbrace{\frac{\sinh(\alpha (b-x_0))}{\sinh(\alpha b)}}_{=\tilde{j}_{\infty,\infty}(0,p|x_0)}  \\  \nonumber
& + \biggl\{\frac{\sinh(\alpha x) \sinh(\alpha x_0)}{\sinh^2(\alpha b)} \frac{\sqrt{\ell_2} I_1(E \sqrt{\ell_1 \ell_2})}{\sqrt{\ell_1}} 
+ \frac{\sinh(\alpha (b-x)) \sinh(\alpha (b-x_0))}{\sinh^2(\alpha b)} \frac{\sqrt{\ell_1} I_1(E \sqrt{\ell_1\ell_2})}{\sqrt{\ell_2}} \\ \label{eq:Pfull}
& + \frac{\sinh(\alpha x_0)\sinh(\alpha (b-x))+\sinh(\alpha x)\sinh(\alpha (b-x_0))}{\sinh^2(\alpha b)}\, I_0(E \sqrt{\ell_1 \ell_2}) \biggr\} 
\frac{E}{2D} e^{-C(\ell_1 + \ell_2)}.
\end{align}
In the opposite case $0 \leq x_0 \leq x \leq b$, one exchanges $x_0$
and $x$.  This is one of the main explicit results of the paper.

\subsection{Probabilistic interpretation}

\begin{figure}
\begin{center}
\includegraphics[width=70mm]{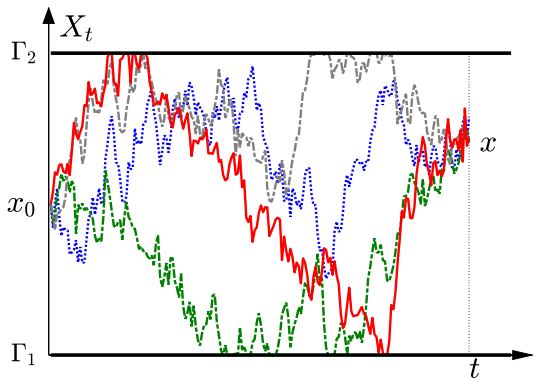} % {traj_int_.eps}
\end{center}
\caption{
Four simulated trajectories on the interval $(0,b)$, each started
from $x_0$ at time $0$ and arrived to $x$ at time $t$: a direct
trajectory that does not hit either of the endpoints (blue dotted
line); a trajectory that does not hit the upper endpoint $\Gamma_2 =
\{b\}$ (green dashed line); a trajectory that does not hit the bottom
endpoint $\Gamma_1 = \{ 0 \}$ (gray dashed line); a trajectory that
hits both endpoints (red solid line).}
\label{fig:traj}
% load('traj_int.mat');
% A_localtime3_fig_traj2(X0,X1,X2,X3);
\end{figure}

Let us discuss the structure of the derived full propagator in
Eq. (\ref{eq:Pfull}).  The first term represents the contributions of
``direct'' trajectories from $x_0$ to $x$ that do not hit either of
the endpoints (Fig. \ref{fig:traj}).  As a consequence, the boundary
local times $\ell_1$ and $\ell_2$ remain zero, as represented by Dirac
distributions $\delta(\ell_1) \delta(\ell_2)$.  The factor in front of
these distributions is the propagator
$\tilde{G}_{\infty,\infty}(x,p|x_0)$ for an interval $(0,b)$ with
absorbing endpoints (i.e., with Dirichlet boundary conditions that
correspond to $q_1 = q_2 = \infty$).  This propagator represents the
fraction of direct trajectories from $x_0$ to $x$.

In the same vein, the second term represents the contribution of
trajectories that do not hit the left endpoint $\Gamma_1$ but may
repeatedly hit the right endpoint $\Gamma_2$.  Introducing the
Laplace-transformed probability flux densities,
\begin{subequations}  \label{eq:jinfty}
\begin{align}
\tilde{j}_{\infty,\infty}(0,p|x_0) & = \left. \bigl(-D \partial_n \tilde{G}_{\infty,\infty}(x,p|x_0)\bigr)\right|_{x=0} 
= \frac{\sinh(\alpha(b-x_0))}{\sinh(\alpha b)} \,, \\
\tilde{j}_{\infty,\infty}(b,p|x_0) & = \left. \bigl(-D \partial_n \tilde{G}_{\infty,\infty}(x,p|x_0)\bigr)\right|_{x=b} 
= \frac{\sinh(\alpha x_0)}{\sinh(\alpha b)} \,,
\end{align}
\end{subequations}
the factor in front of $\delta(\ell_1)$ reads as
\begin{equation*}
\tilde{j}_{\infty,\infty}(b,p|x_0)\, \frac{e^{-C\ell_2}}{D} \, \tilde{j}_{\infty,\infty}(b,p|x). 
\end{equation*}
This factor has a clear probabilistic interpretation: the first
arrival from $x_0$ to $\Gamma_2 = \{b\}$, multiple reflections on that
boundary that increases its boundary local time $\ell_2$ but {\it
conditioned to avoid hitting $\Gamma_1$}, and the last direct move
from $\Gamma_2$ to $x$.  As the computations are performed in Laplace
domain (with respect to time $t$), the product of these three factors
corresponds to the convolution in time domain, as expected.  The
contribution of such multiple returns to $\Gamma_2$ is given by
$e^{-C\ell_2}/D$, where $C = \alpha \, \ctanh(\alpha b)$ can be
interpreted as the eigenvalue of the Dirichlet-to-Neumann operator on
the interval $(0,b)$ with the condition of avoiding $\Gamma_1$ (see
\ref{sec:DN}).

Similarly, the third term with $\delta(\ell_2)$ accounts for the
trajectories that do not hit the right endpoint $\Gamma_2$ but may
repeatedly hit the left one $\Gamma_1$.  The remaining terms in
Eq. (\ref{eq:Pfull}) give the contribution of all other trajectories
that hit both endpoints.

\subsection{Marginal probability quantities}

By integrating the full propagator in Eq. (\ref{eq:Pfull}) over $x$,
we compute the marginal joint probability density of two boundary
local times (in Laplace domain with respect to $t$):
\begin{align}  \nonumber
& \tilde{P}(\circ, \ell_1,\ell_2,p|x_0) =  
\underbrace{\frac{\sinh(\alpha b) - \sinh(\alpha (b-x_0)) - \sinh(\alpha x_0)}{D\alpha^2 \sinh(\alpha b)}}_{=\tilde{S}_{\infty,\infty}(p|x_0)}
 \delta(\ell_1) \delta(\ell_2) \\  \nonumber
& + \frac{\cosh(\alpha b)-1}{\alpha\sinh(\alpha b)} \underbrace{\frac{\sinh(\alpha x_0)}{\sinh(\alpha b)}}
_{=\tilde{j}_{\infty,\infty}(b,p|x_0)} \, \frac{e^{-C\ell_2}}{D} \, \delta(\ell_1)
+ \frac{\cosh(\alpha b)-1}{\alpha \sinh(\alpha b)} \underbrace{\frac{\sinh(\alpha (b-x_0))}{\sinh(\alpha b)}}
_{=\tilde{j}_{\infty,\infty}(0,p|x_0)} \, \frac{e^{-C\ell_1}}{D} \, \delta(\ell_2) \\  \nonumber
& +  \frac{\cosh(\alpha b)-1}{\alpha \sinh(\alpha b)}  \biggl( \frac{\sinh(\alpha x_0)}{\sinh(\alpha b)}
\frac{\sqrt{\ell_2} I_1(E \sqrt{\ell_1 \ell_2})}{\sqrt{\ell_1}} 
+ \frac{\sinh(\alpha (b-x_0))}{\sinh(\alpha b)} \frac{\sqrt{\ell_1} I_1(E \sqrt{\ell_1\ell_2})}{\sqrt{\ell_2}} \\   \label{eq:Pboth}
& + \frac{\sinh(\alpha x_0) + \sinh(\alpha(b-x_0))}{\sinh(\alpha b)} \, I_0(E \sqrt{\ell_1 \ell_2}) \biggr) \frac{E}{2D} e^{-C(\ell_1 + \ell_2)}  .
\end{align}
Its probabilistic interpretation is similar to that of the full
propagator.

In turn, the integral of $\tilde{P}(x,\ell_1,\ell_2,p|x_0)$ over
$\ell_2$ yields the marginal joint probability density of $x$ and
$\ell_1$ (for $0 \leq x \leq x_0 \leq b$):
\begin{align}  \nonumber
\tilde{P}(x,\ell_1,\circ, p|x_0) & = 
\delta(\ell_1) \underbrace{\frac{\sinh (\alpha x)\cosh(\alpha (b-x_0))}{D \alpha \cosh(\alpha b)}}_{=\tilde{G}_{\infty,0}(x,p|x_0)} \\  
\label{eq:interval_xell1}
& + \frac{e^{- \alpha \tanh(\alpha b)\ell_1}}{D} 
\underbrace{\frac{\cosh(\alpha (b- x_0))}{\cosh(\alpha b)}}_{=\tilde{j}_{\infty,0}(0,p|x_0)} 
\underbrace{\frac{\cosh(\alpha (b-x))}{\cosh(\alpha b)}}_{=\tilde{j}_{\infty,0}(0,p|x)}
\end{align}
($x$ and $x_0$ should be exchanged when $x > x_0$).  Expectedly, the
term in front of $\delta(\ell_1)$ is the propagator
$\tilde{G}_{\infty,0}(x,p|x_0)$ for an interval $(0,b)$ with Dirichlet
condition at $x = 0$ and Neumann condition at $x = b$.  In fact, as
one is not interested anymore in the boundary local time $\ell_2$
here, one can put the Neumann boundary condition at $x = b$.  In the
factor $\alpha \tanh(\alpha b)$, one can recognize the eigenvalue of
the Dirichlet-to-Neumann operator on that interval that corresponds to
the eigenfunction $v = 1$ (see \ref{sec:DN}).  Finally, the factor
$\cosh(\alpha(b-x_0))/\cosh(\alpha b)$ is simply the
Laplace-transformed probability flux density
$\tilde{j}_{\infty,0}(0,p|x_0)$.

Integrating Eq. (\ref{eq:interval_xell1}) over $x$, one gets the
marginal probability density of the boundary local time $\ell_1$:
\begin{align}  \nonumber
\tilde{P}(\circ,\ell_1,\circ, p|x_0) & = \delta(\ell_1) 
\underbrace{\frac{\cosh(\alpha b) - \cosh(\alpha (b-x_0))}{D\alpha^2 \cosh(\alpha b)}}_{=\tilde{S}_{\infty,0}(p|x_0)} \\
\label{eq:P_ell1}
& +  \frac{e^{-\alpha \tanh(\alpha b) \ell_1}}{D} \, \underbrace{\frac{\cosh(\alpha (b- x_0))}{\cosh(\alpha b)}}_{=\tilde{j}_{\infty,0}(0,p|x_0)} \, 
\frac{\sinh(\alpha b)}{\alpha \cosh(\alpha b)} \,.
\end{align}
Similarly, the marginal probability density of the boundary local time
$\ell_2$ is
\begin{align}  \label{eq:P_ell2}
\tilde{P}(\circ,\circ,\ell_2,p|x_0) & = \delta(\ell_2) 
\underbrace{\frac{\cosh(\alpha b) - \cosh(\alpha x_0)}{D \alpha^2 \cosh(\alpha b)}}_{=\tilde{S}_{0,\infty}(p|x_0)} + 
\frac{e^{- \alpha \tanh(\alpha b) \ell_2}}{D} \, \underbrace{\frac{\cosh(\alpha x_0)}{\cosh(\alpha b)}}_{=\tilde{j}_{0,\infty}(b,p|x_0)} 
\, \frac{\sinh(\alpha b)}{\alpha \cosh(\alpha b)} \,.
\end{align}

Note also that the joint probability density of the position $X_t$ and
of the total boundary local time, $\ell_t = \ell_t^1 + \ell_t^2$, can
be obtained in the Laplace domain as
\begin{align}  \nonumber
\int\limits_0^\infty d\ell \, e^{-q\ell} \tilde{P}_{\rm tot}(x,\ell,p|x_0) &
= \int\limits_0^\infty d\ell \, e^{-q\ell} \int\limits_0^\infty d\ell_1
\int\limits_0^\infty d\ell_2\, \delta(\ell_1 + \ell_2 - \ell)\, \tilde{P}(x,\ell_1,\ell_2, p|x_0) \\
& = \tilde{G}_{q,q}(x,p|x_0).
\end{align} 
One can either perform the single Laplace transform inversion of
$\tilde{G}_{q,q}(x,p|x_0)$ with respect to $q$, or use the general
spectral expansion derived in \cite{Grebenkov20} based on the
Dirichlet-to-Neumann operator, see Eq. (\ref{eq:Ptot_int}).

Finally, in the limit $b\to \infty$, the full propagator converges to
\begin{equation}  \label{eq:tildeP_binf}
%\fl
\tilde{P}_{b=\infty}(x,\ell_1,\ell_2,p|x_0) = \frac{\delta(\ell_2)}{D} \biggl(e^{-\alpha x_0} \frac{\sinh(\alpha x)}{\alpha} \delta(\ell_1) 
+ e^{-\alpha (x + x_0)} e^{-\alpha \ell_1} \biggr)   \quad  (0 \leq x \leq x_0)
\end{equation}
(in the opposite case $x_0 < x$, one exchanges $x_0$ and $x$).
Expectedly, the boundary local time $\ell_2$ always remains $0$ (see
the factor $\delta(\ell_2)$) as the right endpoint $\Gamma_2$ has
moved to infinity and became unreachable.  Integrating over redundant
variable $\ell_2$, one retrieves thus the full propagator on the
half-line.  Note that the inverse Laplace transform of this expression
can be performed explicitly:
\begin{align}  \nonumber
P_{b=\infty}(x,\ell_1,\circ,t|x_0) & = \delta(\ell_1) \underbrace{\biggl(\frac{\exp\bigl(-\frac{(x-x_0)^2}{4Dt}\bigr)}{\sqrt{4\pi Dt}}
- \frac{\exp\bigl(-\frac{(x+x_0)^2}{4Dt}\bigr)}{\sqrt{4\pi Dt}}\biggr)}_{=G_{\infty}(x,t|x_0)}  \\
& + (x+x_0+\ell_1) \frac{\exp\bigl(-\frac{(x+x_0+\ell_1)^2}{4Dt}\bigr)}{\sqrt{4\pi D^3 t^3}} \,.
\end{align}
In turn, the integral of Eq. (\ref{eq:tildeP_binf}) over $x$ yields
the marginal probability density of $\ell_1$:
\begin{equation}
\tilde{P}_{b=\infty}(\circ,\ell_1,\circ,p|x_0) = \frac{1-e^{-\alpha x_0}}{D\alpha^2} \delta(\ell_1) 
+ \frac{e^{- \alpha x_0}}{D\alpha} e^{-\alpha \ell_1} \qquad (x_0\geq 0) ,
\end{equation}
which can also be inverted:
\begin{equation}
P_{b=\infty}(\circ,\ell_1,\circ,t|x_0) = \erf\biggl(\frac{x_0}{\sqrt{4Dt}}\biggr) \delta(\ell_1) 
+ \frac{\exp\bigl(-\frac{(x_0 + \ell_1)^2}{4Dt}\bigr)}{\sqrt{\pi Dt}}  \quad (x_0 \geq 0) ,
\end{equation}
in agreement with Ref. \cite{Grebenkov19g}.

\subsection{Joint cumulative probability function}

The statistics of two boundary local times $\ell_t^1$ and $\ell_t^2$
is fully determined by the marginal joint probability density
$P(\circ,\ell_1,\ell_2,t|x_0)$.  For some applications (see below), it
is more convenient to deal with the joint cumulative probability
function:
\begin{equation}  \label{eq:F_def}
F(\ell_1,\ell_2,t|x_0) = \int\limits_0^{\ell_1} d\ell'_1 \int\limits_0^{\ell_2} d\ell'_2 \, P(\circ,\ell'_1,\ell'_2,t|x_0) .
\end{equation}
Using Eq. (\ref{eq:Pboth}), we obtain after simplifications
\begin{align}  \nonumber
\tilde{F}(\ell_1,\ell_2,p|x_0) & = \tilde{S}_{\infty,\infty}(p|x_0) 
+ \frac{\sinh(\alpha x_0) + \sinh(\alpha(b-x_0))}{D\alpha^2 \sinh(\alpha b)} Q_2\bigl(C\ell_1,C\ell_2; \sech(\alpha b)\bigr) \\  \nonumber
& + \frac{(\cosh(\alpha b)-1) \sinh(\alpha x_0)}{D\alpha^2  \cosh(\alpha b) \sinh(\alpha b) } 
e^{- \alpha \tanh(\alpha b) \ell_1} Q_1\bigl(C\ell_2; \sqrt{C\ell_1} \, \sech(\alpha b)\bigr)  \\  \label{eq:tildeF}
& + \frac{(\cosh(\alpha b)-1) \sinh(\alpha (b-x_0))}{D\alpha^2 \cosh(\alpha b) \sinh(\alpha b)} 
e^{- \alpha \tanh(\alpha b) \ell_2} Q_1\bigl(C\ell_1; \sqrt{C\ell_2} \, \sech(\alpha b)\bigr) ,
\end{align}
where $\sech(z) = 1/\cosh(z)$, 
\begin{equation}
\tilde{S}_{\infty,\infty}(p|x_0) = \frac{\sinh(\alpha b) - \sinh(\alpha (b-x_0)) - \sinh(\alpha x_0)}{D\alpha^2 \sinh(\alpha b)} \,,
\end{equation}
and we introduced two auxiliary functions:
\begin{subequations} \label{eq:Q1Q2}
\begin{align}
Q_1(z;a) & = e^{-a^2} \int\limits_0^z dx \, e^{-x} \, I_0(2a\sqrt{x}) , \\ 
Q_2(z_1,z_2;a) & = (1-a^2) \int\limits_0^{z_1} dx_1 \int\limits_0^{z_2} dx_2 \, e^{-x_1-x_2} \, I_0(2a\sqrt{x_1x_2}) .
\end{align}
\end{subequations}
Strictly speaking, the first term in Eq. (\ref{eq:tildeF}) should
include the Heaviside functions $\Theta(\ell_1) \Theta(\ell_2)$, which
after differentiation with respect to $\ell_1$ and $\ell_2$ yields
$\delta(\ell_1) \delta(\ell_2)$ in the first term in
Eq. (\ref{eq:Pfull}).  Similarly, some other terms should include
$\Theta(\ell_1)$ and $\Theta(\ell_2)$ but we omit them for brevity by
considering $\ell_1 > 0$ and $\ell_2 > 0$.
Setting $\ell_1 = \ell_2 = 0$, one retrieves the Laplace-transformed
survival probability $\tilde{S}_{\infty,\infty}(p|x_0)$, as expected.

The definition of the functions $Q_1$ and $Q_2$ ensures that
$Q_1(\infty;a) = 1$ and $Q_2(\infty,\infty;a) = 1$ so that
$\tilde{F}(\infty,\infty,p|x_0) = 1/p$ and thus
$F(\infty,\infty,t|x_0) = 1$ as expected.  Moreover, since
$Q_1(z;\infty) = 0$ and $Q_2(\infty,z;a) = 1 - e^{-z(1-a^2)}$, one
easily finds the Laplace transforms of the marginal cumulative
probability functions:
\begin{subequations}  \label{eq:tildeF_marg}
\begin{align}
\tilde{F}(\ell_1,\infty,p|x_0) & = \frac{1}{D\alpha^2} \biggl(1 - 
\frac{\cosh(\alpha (b-x_0))}{\cosh(\alpha b)} e^{-\alpha \tanh(\alpha b) \ell_1} \biggr) , \\
\tilde{F}(\infty,\ell_2,p|x_0) & = \frac{1}{D\alpha^2} \biggl(1 - 
\frac{\cosh(\alpha x_0)}{\cosh(\alpha b)} e^{-\alpha \tanh(\alpha b) \ell_2} \biggr)  
\end{align}
\end{subequations}
(see \ref{sec:Q1Q2} for some other properties of the functions $Q_1$
and $Q_2$).

\subsection{Results in time domain}
\label{sec:time}

The above expressions determine the full propagator and marginal
densities in Laplace domain with respect to time $t$.  As it is quite
common for diffusion-based quantities, representations in Laplace
domain are more explicit and compact.  A standard way to perform the
Laplace inversion and thus to pass back to time domain consists in
searching for the poles of the full propagator in the complex plane
$p\in\C$.  For instance, this computation is straightforward for the
first term in Eq. (\ref{eq:Pfull}) and yields the standard spectral
expansion of the propagator $G_{\infty,\infty}(x,t|x_0)$ on the
interval with absorbing endpoints.  However, the analysis is more
subtle for other terms.  For example, the second term in front of
$\delta(\ell_1)$ includes the function
\begin{equation*}
\tilde{f}(p) = e^{-C\ell_2} = \exp\biggl(-\ell_2 \sqrt{p/D} \, \ctanh\bigl(\sqrt{p/D}\, b\bigr) \biggr).  
\end{equation*}
While the poles of the factor in front of this exponential function
are $p_n = - \pi^2 n^2 D/b^2$, the function $\tilde{f}(p)$ rapidly
vanishes as $p\to p_n$ that prevents applying the residue theorem.  In
\ref{sec:Laplace_inv}, we derive a semi-analytical formula
for inverting such Laplace transforms.  This formula is particularly
valuable in the short-time limit.  However, its practical
implementation becomes numerically difficult at long times.
For this reason, we applied the Talbot algorithm for numerical Laplace
transform inversion.  Further analysis of the long-time asymptotic
behavior remains an interesting open problem.

\begin{figure}
\begin{center}
\includegraphics[width=160mm]{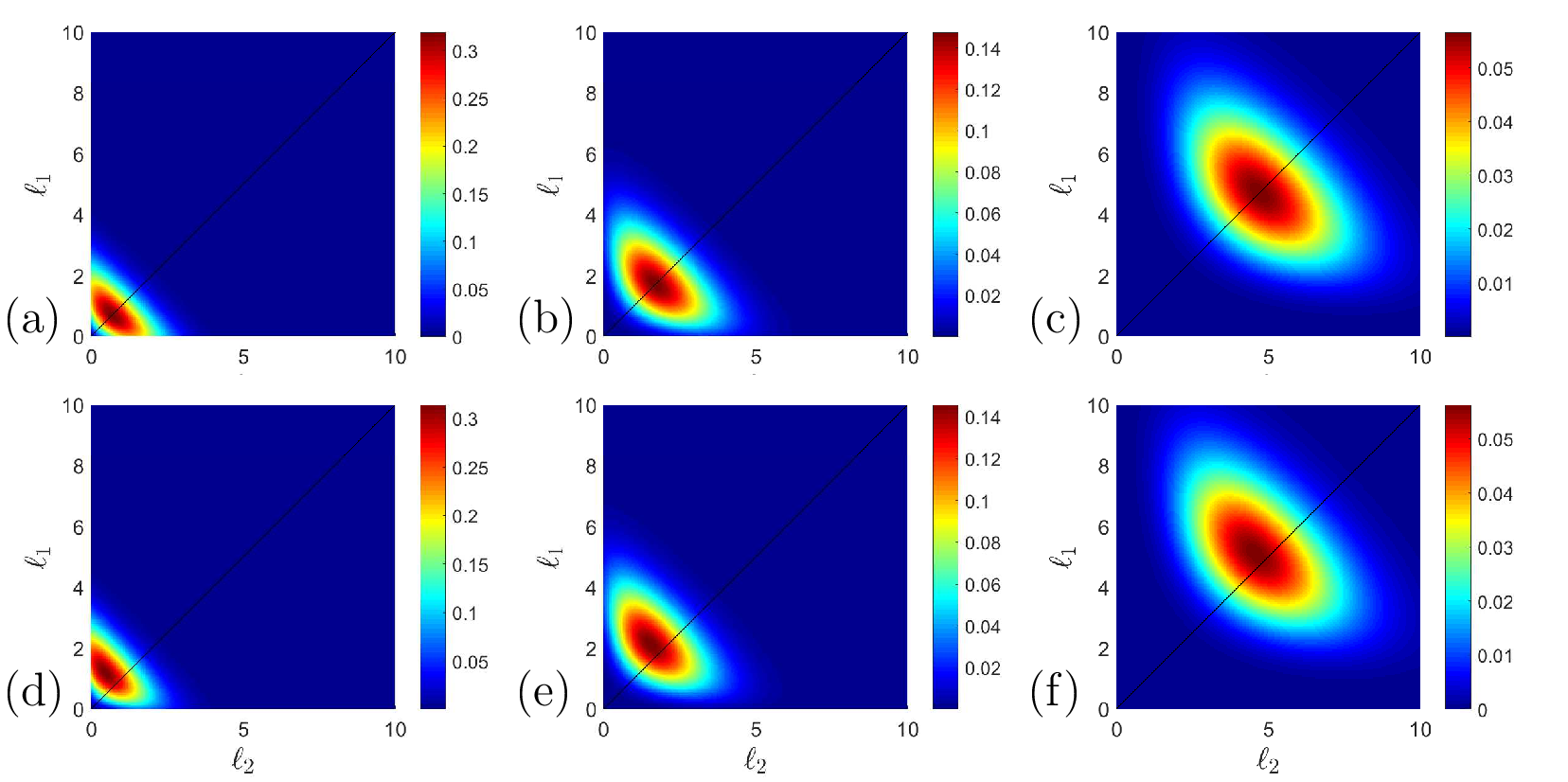}
\end{center}
\caption{
The continuous part of the joint probability density
$P(\circ,\ell_1,\ell_2,t|x_0)$ of two boundary local times $\ell_1$
and $\ell_2$ at the endpoints of the unit interval $(0,1)$, with $x_0
= 0.5$ (top row) and $x_0 = 0$ (bottom row), and three times: $t = 1$
(left column), $t = 2$ (middle column), and $t = 5$ (right column),
with $D = 1$.  }
\label{fig:rhot}
% save('rhot_x0_01.mat');
% A_localtime3_joint_t_fig(rhot_t1);
\end{figure}

Figure \ref{fig:rhot} shows the joint probability density
$P(\circ,\ell_1,\ell_2,t|x_0)$ of two boundary local times $\ell_1$
and $\ell_2$ at the endpoints of the unit interval $(0,1)$.  Here we
present only the continuous part (i.e., the three last terms that do
not contain either $\delta(\ell_1)$, nor $\delta(\ell_2)$; in fact,
the three other terms containing either of these $\delta$'s are
simpler and can be presented separately).  When the starting point
$x_0$ is at the middle of the interval (top row), both endpoints are
equally accessible, and $P(\circ,\ell_1,\ell_2,t|x_0)$ is symmetric
with respect to exchange of $\ell_1$ and $\ell_2$.  As time $t$
increases, the maximum of the joint probability density moves along
the diagonal $\ell_1 = \ell_2$.  In fact, in the long-time limit
($\sqrt{Dt} \gg b$), the diffusing particle has enough time to
frequently encounter both endpoints, and the mean boundary local times
grow linearly with $t$, see Eq. (\ref{eq:mean_tinf}).  As the variance
also grows linearly with time according to Eq. (\ref{eq:var_ell}), the
maximum of the joint probability density spreads.
If the particle starts on (or near) the left endpoint (bottom row),
the joint probability density is shifted to larger values of $\ell_1$.
However, as time increases, the maximum progressively returns to the
diagonal, as expected.

\subsection{Extension to circular annulus and spherical shell}
\label{sec:extensions}

The computation for a circular annulus, $\Omega = \{ \x\in\R^2 ~:~ a <
|\x| < b\}$, and for a spherical shell, $\Omega = \{ \x\in\R^3 ~:~ a <
|\x| < b\}$, are very similar but technically more involved.  In fact,
the rotational symmetry of these domains allows one to separate
variables and to expand the solution over Fourier harmonics (in 2D) or
over spherical harmonics (in 3D).  In turn, the radial propagator
associated to each harmonic has an exact explicit form, which is
similar to Eq. (\ref{eq:tildeG}), see \cite{Grebenkov19g} and
\ref{sec:2d_3d} for details.  As the dependence on $q_1$ and $q_2$ is
exactly the same, one can apply the above technique to inverse the
double Laplace transform.  One gets then the radial part of the full
propagator, $\tilde{P}_n(r,\ell_1,\ell_2,p|r_0)$ (corresponding to the
$n$-th harmonic).  The structure of this radial part is similar to
that of Eq. (\ref{eq:Pfull}), even though the radius-dependent
prefactors are different.  For the sake of brevity, we do not provide
explicit formulas here (see also \cite{Grebenkov20b}).

\section{Variety of first-passage times}
\label{sec:FPT}

The derived joint probability densities allow one to investigate
various first-passage times (FPTs).  The distribution of a
first-passage time $\tau$ is in general determined by the survival
probability, $\P_{x_0}\{ \tau > t\}$, from which the probability
density follows as $H(t|x_0) = -\partial_t \P_{x_0}\{ \tau > t\}$.  We
will consider the latter quantity in the Laplace domain.

\subsection{Conventional first-passage times}

The distribution of the FPT to a perfectly or partially reactive
target has been intensively studied in various settings
\cite{Redner,Metzler,Oshanin,Bray13,Benichou08,Benichou10,Rupprecht15,Godec16a,Godec16b,Grebenkov18,Agranov18,Lanoiselee18,Artime18,Grebenkov18d,Levernier19,Grebenkov19e,Grebenkov19f,Lawley20,Grebenkov20a}.
The most common first-passage time is the moment of the first arrival
of the process to the boundary (or the target): $\tau =
\inf\{ t>0 ~:~ \X_t \in \pa\}$.  As the boundary local time remains
zero until the first encounter, this first-passage time can also be
formulated as $\tau = \inf\{ t > 0 ~:~ \ell_t > 0\}$, i.e., the moment
of the first crossing of the threshold $0$ by the total boundary local
time $\ell_t$.
In the case of the interval, the FPT to either of the boundaries
$\Gamma_1$ and $\Gamma_2$ reads then
\begin{equation}
%\fl \qquad 
\tau = \tau_{\infty,\infty} = \inf\{t > 0 ~:~ \ell_t^1 + \ell_t^2 > 0\} = \inf\{t > 0 ~:~ \ell_t^1 > 0 ~ \textrm{or} ~ \ell_t^2 > 0\} .
\end{equation}
This FPT is determined by the Laplace-transformed survival probability
$\tilde{S}_{\infty,\infty}(p|x_0)$ standing in front of
$\delta(\ell_1) \delta(\ell_2)$ in Eq. (\ref{eq:Pboth}):
\begin{equation}
\P_{x_0}\{ \tau_{\infty,\infty} > t\} = \P_{x_0}\{ \ell_t^1 = 0 ~ \textrm{and}~ \ell_t^2 = 0\} = S_{\infty,\infty}(t|x_0).
\end{equation}

Similarly, one can consider the FPT to one endpoint, say, to
$\Gamma_2$: $\tau_{0,\infty} = \inf\{t > 0 ~:~ \ell_t^2 > 0\}$.  The
condition $\ell_t^2 = 0$ is expressed by $\delta(\ell_2)$, which is
present in the first and the third terms in Eq. (\ref{eq:Pboth}).
Integrating these terms over the marginal variable $\ell_1$ from $0$
to $\infty$, one gets
\begin{align} \nonumber
& \P_{x_0}\{ \tau_{0,\infty} > t\} = \P_{x_0}\{ \ell_t^2 = 0\} \\  \nonumber
& = \int\limits_0^\infty d\ell_1 \biggl(
\tilde{S}_{\infty,\infty}(p|x_0) \, \delta(\ell_1)   
+  \frac{\cosh(\alpha b)-1}{\alpha \sinh(\alpha b)} \frac{\sinh(\alpha (b-x_0))}{\sinh(\alpha b)} \, \frac{e^{-C\ell_1}}{D} \biggr)  
= \tilde{S}_{\infty,\infty}(p|x_0) \\
& + \frac{(\cosh(\alpha b)-1)}{D\alpha^2\cosh(\alpha b)}  \, \frac{\sinh(\alpha (b-x_0))}{\sinh(\alpha b)}
= \frac{1}{D\alpha^2} \biggl(1 - \frac{\cosh(\alpha x_0)}{\cosh(\alpha b)}\biggr) = \tilde{S}_{0,\infty}(p|x_0).
\end{align}
Indeed, as we are not interested in the boundary local time $\ell_t^1$
here, this is equivalent to setting Neumann boundary condition on
$\Gamma_1$, as discussed above.

\subsection{First reaction times}

When both endpoints are partially absorbing with equal reactivities
(i.e., $q_1 = q_2 = q$), the reaction time can be defined as
$\tau_{q,q} = \inf\{ t>0 ~:~ \ell_t^1 + \ell_t^2 > \hat{\ell}\}$,
i.e., the first moment when the total boundary local time exceeds a
random independently distributed threshold $\hat{\ell}$ with the
exponential distribution with the mean $q$: $\P\{\hat{\ell} > \ell
\} = e^{-q\ell}$ \cite{Grebenkov06,Grebenkov07a,Grebenkov19b,Grebenkov20}.
Qualitatively, the exponentially distributed threshold $\hat{\ell}$
for surface reactions plays the same role as an exponentially
distributed lifetime of a particle for bulk reactions (see
\cite{Grebenkov20} for details).  The distribution of this random
reaction time is 
\begin{equation}
\P_{x_0}\{\tau_{q,q} > t\} = S_{q,q}(t|x_0), 
\end{equation}
which is determined by the explicitly known $\tilde{S}_{q,q}(t|x_0)$
from Eq. (\ref{eq:tildeS}).  This is a common setting for partial
reactivity.

The above setting can be naturally generalized to deal with distinct
surface reactivity parameters $q_1$ and $q_2$.  In this case, one has
to consider two boundary local times separately, as encounters with
$\Gamma_1$ and $\Gamma_2$ result in the reaction event in different
ways.  Here, we define
\begin{equation}  \label{eq:tau_q1q2}
\tau_{q_1,q_2} = \inf\{ t>0 ~:~ \ell_t^1 > \hat{\ell}_1  ~ \textrm{or}  ~ \ell_t^2 > \hat{\ell}_2\} ,
\end{equation}
as the first moment when {\it either} of the boundary local times
$\ell_t^1$ and $\ell_t^2$ exceeds its random threshold, $\hat{\ell}_1$
and $\hat{\ell}_2$, which are determined as independent exponential
random variables with means $q_1$ and $q_2$, respectively.  As
boundary local times are nondecreasing processes, the event $\{
\tau_{q_1,q_2} > t\}$ means that none of boundary local times exceeded
its threshold:
\begin{align}  \nonumber
\P_{x_0}\{\tau_{q_1,q_2} > t\} & = \P_{x_0}\bigl\{ \ell_t^1 < \hat{\ell}_1 ~ \textrm{and} ~ \ell_t^2 < \hat{\ell}_2\bigr\} \\  \nonumber
& = \int\limits_0^{\infty} d\ell_1 \int\limits_0^{\infty} d\ell_2 \, P(\circ,\ell_1,\ell_2,t|x_0) \, 
\P\{ \ell_1 < \hat{\ell}_1~ \textrm{and} ~ \ell_2 < \hat{\ell}_2 \} \\
& = \int\limits_0^{\infty} d\ell_1 \int\limits_0^{\infty} d\ell_2 \, P(\circ,\ell_1,\ell_2,t|x_0) \, e^{-q_1\ell_1} \, e^{-q_2 \ell_2} 
= S_{q_1,q_2}(t|x_0) ,
\end{align}
where we applied Eq. (\ref{eq:Srho_general}) and used that
$\hat{\ell}_1$ and $\hat{\ell}_2$ are independent exponential
variables.  In other words, this FPT time is determined by the
survival probability $S_{q_1,q_2}(t|x_0)$ with Robin boundary
conditions (\ref{eq:S_Robin}), as expected.  While this extension is
natural, we are not aware of earlier probabilistic definitions of the
FPT $\tau_{q_1,q_2}$ with the help of two boundary local times, as in
Eq. (\ref{eq:tau_q1q2}).

\subsection{First-crossing times of two thresholds}

The explicit form of the joint probability density $\tilde{P}(\circ,
\ell_1,\ell_2,p|x_0)$ allows one to go far beyond the aforementioned
first-passage times.  In particular, we will generalize the
probability density of the first-crossing time for the total boundary
local time $\ell_t$ derived in \cite{Grebenkov20} (see
\ref{sec:A_FPT} for its properties).

\subsubsection*{First crossing by either of two boundary local times}

The first natural extension consists in replacing exponential
thresholds $\hat{\ell}_1$ and $\hat{\ell}_2$ in
Eq. (\ref{eq:tau_q1q2}) by fixed thresholds $\ell_1$ and $\ell_2$.  In
other words, we are interested in the first moment when either of two
boundary local times exceeds its threshold:
\begin{equation}
\tau_{\cup} = \inf\{ t>0 ~:~ \ell_t^1 > \ell_1  ~ \textrm{or}  ~ \ell_t^2 > \ell_2\} .
\end{equation}
For instance, this FPT can describe the moment of the reaction, which
is initiated when the particle either has visited at least $\ell_1/a$
times the vicinity of width $a$ of the left target, or has visited at
least $\ell_2/a$ times the $a$-vicinity of the right target.
Qualitatively, this FPT describes a sort of minimal condition to
produce the reaction event by either of the targets.  If $\ell_1 =
\ell_2 = \ell$, then $\tau_{\cup}$ is the first moment when
$\max\{\ell_t^1, \ell_t^2\}$ exceeds $\ell$.

\begin{figure}
\begin{center}
\includegraphics[width=70mm]{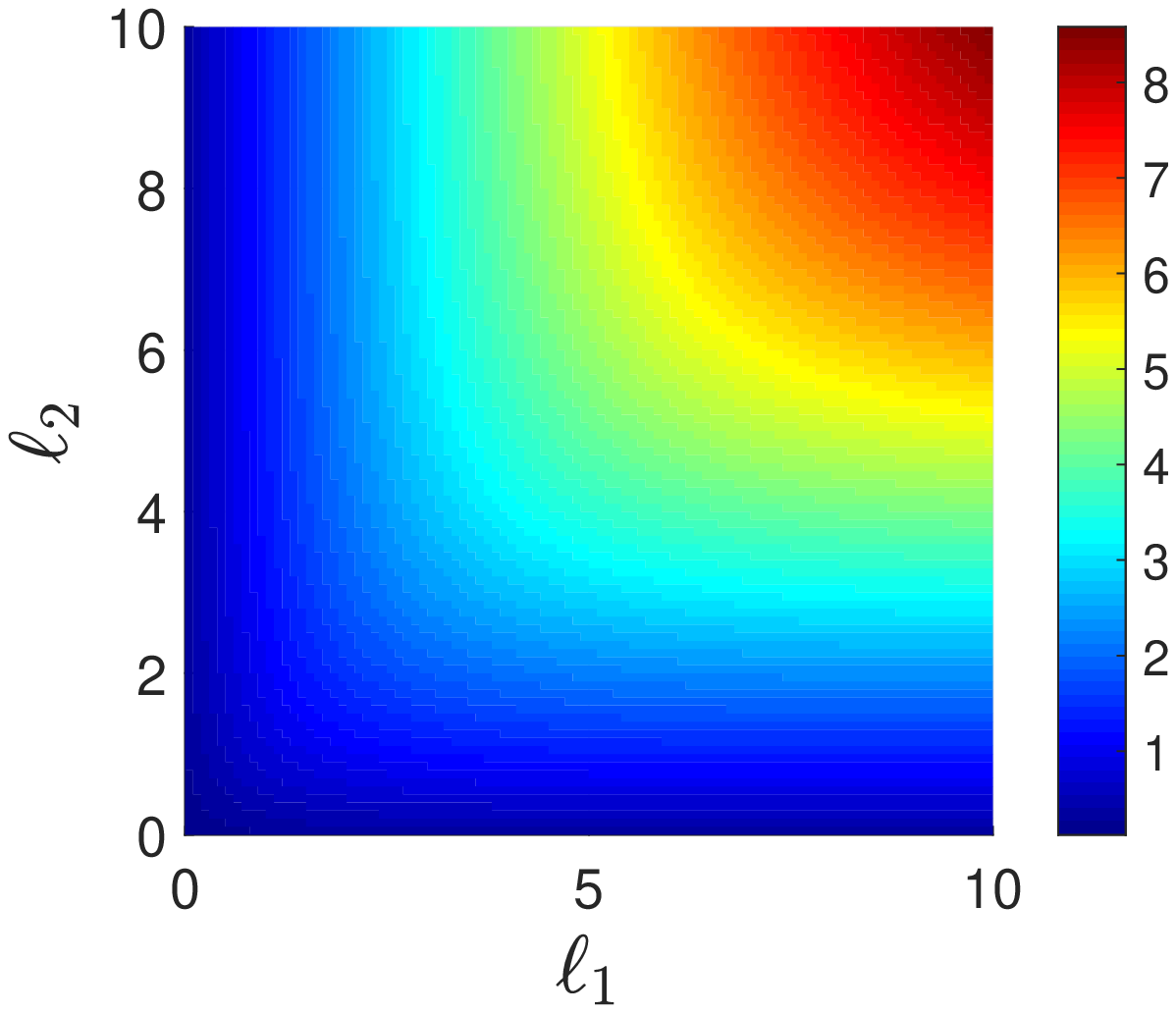} % {Tmean_cup.eps}
\includegraphics[width=70mm]{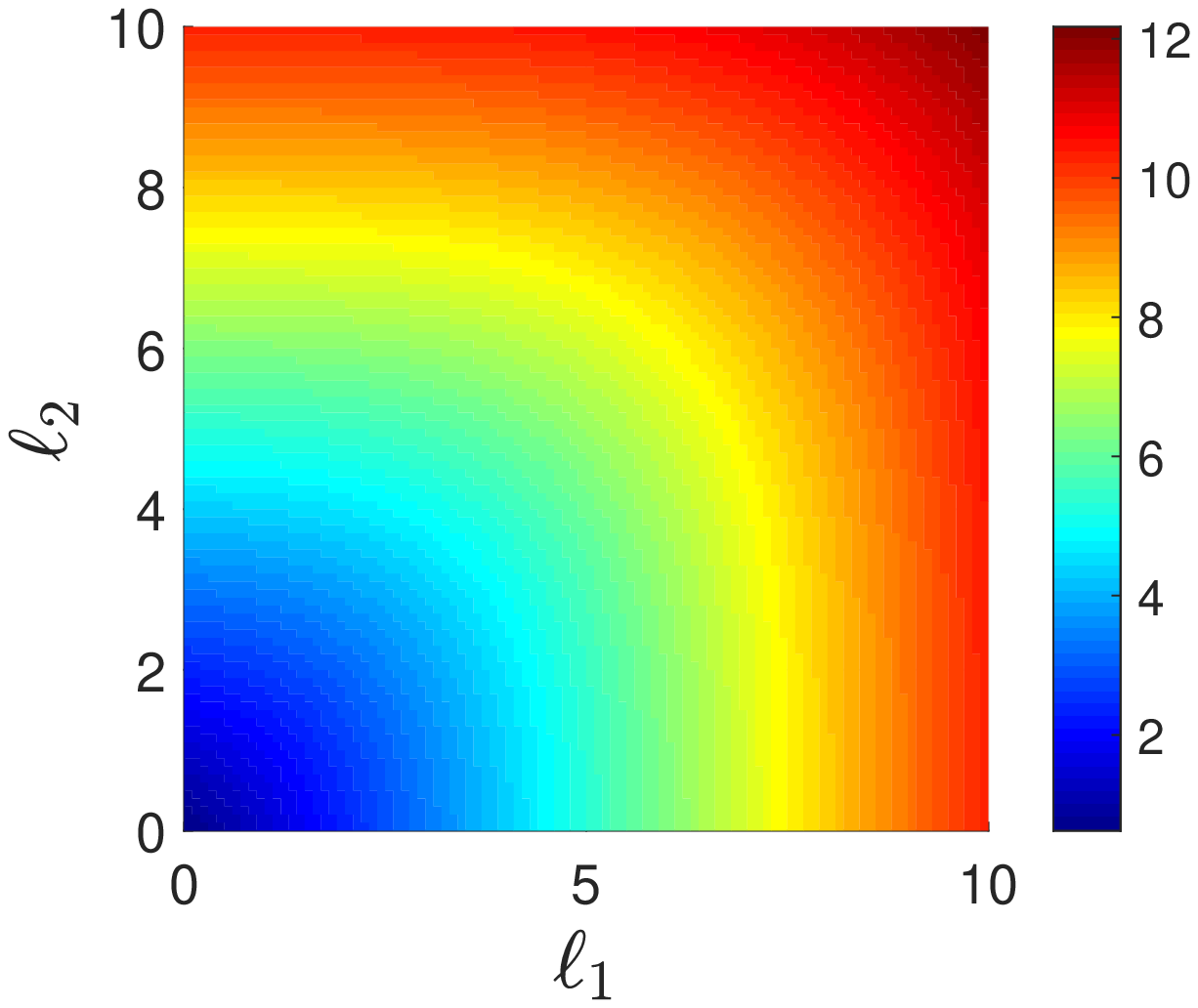} % {Tmean_cap.eps}
\end{center}
\caption{
The mean first-crossing times $\E_{x_0}\{ \tau_{\cup} \}$ (left) and
$\E_{x_0}\{ \tau_{\cap}\}$ (right) as functions of thresholds $\ell_1$
and $\ell_2$, with $b = 1$, $D = 1$, and $x_0 = 0.5$.}
\label{fig:Tmean}
% [Tcup,Tcap] = A_localtime3_mean_tau_fig(Tcup,Tcap,0);
% [Tcup,Tcap] = A_localtime3_mean_tau_fig(Tcup,Tcap,1);
\end{figure}

The first-crossing time $\tau_{\cup}$ is determined by
\begin{align}
S_{\cup}(t|x_0) & = \P_{x_0}\{\tau_{\cup} > t\} = \P_{x_0}\bigl\{ \ell_t^1 < \ell_1 ~ \textrm{and} ~ \ell_t^2 < \ell_2\bigr\} 
= F(\ell_1,\ell_2,t|x_0),
\end{align}
where $F(\ell_1,\ell_2,t|x_0)$ is the joint cumulative probability
function defined in Eq. (\ref{eq:F_def}).  As the probability density
of the first-crossing time, $H_{\cup}(t|x_0)$, is obtained by taking
the time derivative of the survival probability (with negative sign),
we get in the Laplace domain:
\begin{equation}  \label{eq:H_cup}
\tilde{H}_{\cup}(p|x_0) = 1 - p \, \tilde{F}(\ell_1,\ell_2,p|x_0),
\end{equation}
with $\tilde{F}(\ell_1,\ell_2,p|x_0)$ given by Eq. (\ref{eq:tildeF}).
As usual, this function determines all positive integer moments of
$\tau_{\cup}$:
\begin{equation}
\E_{x_0}\{ \tau_{\cup}^m \} = (-1)^m \lim\limits_{p\to 0} \frac{\partial^m}{\partial p^m} \tilde{H}_{\cup}(p|x_0).
\end{equation}
In particular, the mean first-crossing time is
\begin{align}  \nonumber
\E_{x_0}\{ \tau_{\cup} \} & = \tilde{F}(\ell_1,\ell_2, 0|x_0) 
= \frac{x_0(b-x_0)}{2D} + \frac{b^2}{D} \hat{Q}_2(\ell_1/b,\ell_2/b) \\ \label{eq:mean_tau}
& + \frac{bx_0}{2D} Q_1(\ell_2/b;\sqrt{\ell_1/b}) + \frac{b(b-x_0)}{2D} Q_1(\ell_1/b;\sqrt{\ell_2/b}) ,
\end{align}
where
\begin{equation}
\hat{Q}_2(z_1,z_2) = \lim\limits_{a\to 1} \frac{Q_2(z_1,z_2;a)}{1-a^2} = \int\limits_0^{z_1} dx_1 
\int\limits_0^{z_2} dx_2 \, e^{-x_1-x_2} I_0(2\sqrt{x_1x_2}) .
\end{equation}
Figure \ref{fig:Tmean}(left) illustrates the behavior of the mean
first-crossing time $\E_{x_0}\{ \tau_{\cup} \}$ as a function of
$\ell_1$ and $\ell_2$.

As discussed in Sec. \ref{sec:time}, the analytical inversion of the
Laplace transform like that in Eq. (\ref{eq:H_cup}) is a challenging
task.  However, the short-time asymptotic behavior of the probability
density can be easily obtained.  Using the asymptotic relations from
\ref{sec:Q1Q2}, we get in the limit $p\to \infty$ for any $0
< x_0 < b$:
\begin{equation}
\tilde{F}(\ell_1,\ell_2,p|x_0) \simeq \frac{1}{p}\biggl(1 - e^{-\alpha (x_0 + \ell_1)} - e^{-\alpha (b-x_0+\ell_2)} \biggr).
\end{equation}
The short-time behavior of the probability density $H_{\cup}(t|x_0)$
follows then
\begin{equation}  \label{eq:Hcup_short}
%\fl \quad
H_{\cup}(t|x_0) \simeq \frac{1}{\sqrt{4\pi D t^3}} \biggl( (x_0+\ell_1) e^{-(x_0+\ell_1)^2/(4Dt)} 
+ (b-x_0+\ell_2)e^{-(b-x_0+\ell_2)^2/(4Dt)} \biggr).
\end{equation}
Qualitatively, the first term represents the contribution of
trajectories that rapidly reached the left endpoint (by crossing the
distance $x_0$) and remained in its vicinity to increase the boundary
local time $\ell_t^1$ up to $\ell_1$.  Similarly, the second term
accounts for the trajectories that reached the right endpoint and
stayed nearby.

\begin{figure}
\begin{center}
\includegraphics[width=100mm]{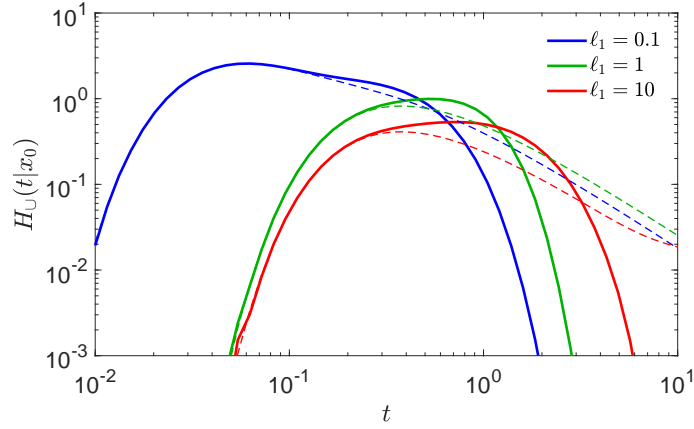} % {Hcup.eps}
\end{center}
\caption{
The probability density $H_{\cup}(t|x_0)$ of the first-crossing time
$\tau_{\cup}$, with $b = 1$, $D = 1$, $x_0 = 0.5$, $\ell_2 = 1$, and
three values of $\ell_1$ as indicated in the plot.  Solid lines show
the numerical inversion of $\tilde{H}_\cup(p|x_0)$ via the Talbot
algorithm, whereas dashed lines indicate the short-time asymptotic
relation (\ref{eq:Hcup_short}).}
\label{fig:Ht_cup}
% A_localtime3_FH_t_fig2(rhot1,rhot2,rhot3);
\end{figure}

Figure \ref{fig:Ht_cup} presents three probability densities
$H_{\cup}(t|x_0)$ for $\ell_2 = 1$ and three values of $\ell_1$:
$0.1$, $1$, and $10$.  One first notes that the short-time relation
(\ref{eq:Hcup_short}) is in excellent agreement with the numerical
inversion of $\tilde{H}_\cup(p|x_0)$ via the Talbot algorithm.  As
$\tau_\cup$ characterizes the first moment when either of two boundary
local times crosses its threshold, the density $H_{\cup}(t|x_0)$ is
shifted toward shorter times for $\ell_1 = 0.1$.  In fact, it is on
average much faster for the boundary local time $\ell_t^1$ to cross
the threshold $\ell_1 = 0.1$ than for $\ell_t^2$ to cross $\ell_2 =
1$.  The opposite situation occurs for $\ell_1 = 10$, which takes
longer to cross than $\ell_2 = 1$.  This explains that the probability
density $H_{\cup}(t|x_0)$ does not considerably change when $\ell_1$
is increased from $1$ to $10$.

While the short-time behavior is available, getting the long-time
asymptotic behavior of $H_{\cup}(t|x_0)$ remains an open problem (see
the related discussion in \ref{sec:A_FPT} for a similar problem in the
case of the total boundary local time).

\subsubsection*{First crossing by both boundary local times}

Alternatively, one can look at the first moment when both $\ell_t^1$
and $\ell_t^2$ exceed their thresholds:
\begin{equation}
\tau_{\cap} = \inf\{ t>0 ~:~ \ell_t^1 > \ell_1  ~ \textrm{and}  ~ \ell_t^2 > \ell_2\} .
\end{equation}
In particular, if $\ell_1 = \ell_2 = \ell$, $\tau_{\cap}$ is the first
moment when $\min\{\ell_t^1, \ell_t^2\}$ exceeds $\ell$.

The first-crossing time $\tau_{\cap}$ is determined by
\begin{align}  \nonumber
S_{\cap}(t|x_0) & = \P_{x_0}\{\tau_{\cap} > t\} = \P_{x_0}\bigl\{ \ell_t^1 < \ell_1 ~ \textrm{or} ~ \ell_t^2 < \ell_2\bigr\} \\  \nonumber
& = 1 - \P_{x_0}\bigl\{ \ell_t^1 > \ell_1 ~ \textrm{and} ~ \ell_t^2 > \ell_2\bigr\} 
= 1 - \int\limits_{\ell_1}^\infty d\ell'_1 \int\limits_{\ell_2}^\infty d\ell'_2 \, P(\circ,\ell'_1,\ell'_2,t|x_0) \\
& = F(\ell_1,\infty, t|x_0) + F(\infty,\ell_2,t|x_0) - F(\ell_1,\ell_2,t|x_0),
\end{align}
where the first two terms correspond to marginal cumulative
probability functions given by Eqs. (\ref{eq:tildeF_marg}).  In the
Laplace domain, we get then
\begin{align}  \nonumber
\tilde{H}_{\cap}(p|x_0) & = 1 - p \, \biggl(\tilde{F}(\ell_1,\infty,p|x_0) + \tilde{F}(\infty,\ell_2,p|x_0) - \tilde{F}(\ell_1,\ell_2,p|x_0) \biggr) \\
\label{eq:H_cap}
& = \frac{\cosh(\alpha (b-x_0))}{\cosh (\alpha b)} e^{-\alpha \tanh(\alpha b) \ell_1} 
+ \frac{\cosh(\alpha x_0)}{\cosh (\alpha b)} e^{-\alpha \tanh(\alpha b) \ell_2}  - \tilde{H}_{\cup}(p|x_0) ,
\end{align}
where $\tilde{H}_{\cup}(p|x_0)$ is given by Eq. (\ref{eq:H_cup}), and
we used Eqs. (\ref{eq:tildeF_marg}).

As previously, the density $\tilde{H}_{\cap}(p|x_0)$ determines all
the positive integer moments of $\tau_{\cap}$, in particular,
\begin{align}  \nonumber
\E_{x_0}\{ \tau_{\cap} \} & = \tilde{F}(\ell_1,\infty,0|x_0) + \tilde{F}(\infty,\ell_2,0|x_0) - \tilde{F}(\ell_1,\ell_2,0|x_0) \\
& = \frac{2b(\ell_1 + \ell_2) + 2b^2 - x_0^2 - (b-x_0)^2}{2D} - \E_{x_0}\{  \tau_{\cup}\} ,
\end{align}
where $\E_{x_0}\{ \tau_{\cup} \}$ is given by Eq. (\ref{eq:mean_tau}).
Figure \ref{fig:Tmean}(right) illustrates the behavior of the mean
first-crossing time $\E_{x_0}\{ \tau_{\cap} \}$ as a function of
$\ell_1$ and $\ell_2$.

The short-time asymptotic behavior is determined from the limit $p\to
\infty$.  In this case, the leading terms that determined the behavior
of $H_\cup(t|x_0)$, vanish, and one needs to keep terms up to the
order of $e^{-\alpha b}$.  Skipping technical details, we get
\begin{align}  \nonumber
H_{\cap}(t|x_0) & \simeq \frac{1}{\sqrt{\pi D t^3}} \biggl( (x_0+\ell_1+\ell_2+b) e^{-(x_0+\ell_1+\ell_2+b)^2/(4Dt)}  \\  \label{eq:Hcap_short}
& + (2b-x_0+\ell_1+\ell_2)e^{-(2b-x_0+\ell_1+\ell_2)^2/(4Dt)} \biggr).
\end{align}
Qualitatively, the first term represents the contribution of
trajectories that rapidly reached the left endpoint (by crossing the
distance $x_0$) and remained in its vicinity to increase the boundary
local time $\ell_t^1$ up to $\ell_1$, then crossed the interval (by
traveling distance $b$) to reach the right endpoint and remained
nearby to increase $\ell_t^2$ up to $\ell_2$.  Similarly, the second
term accounts for the trajectories that first reached the right
endpoint and then moved to the left endpoint.

\begin{figure}
\begin{center}
\includegraphics[width=100mm]{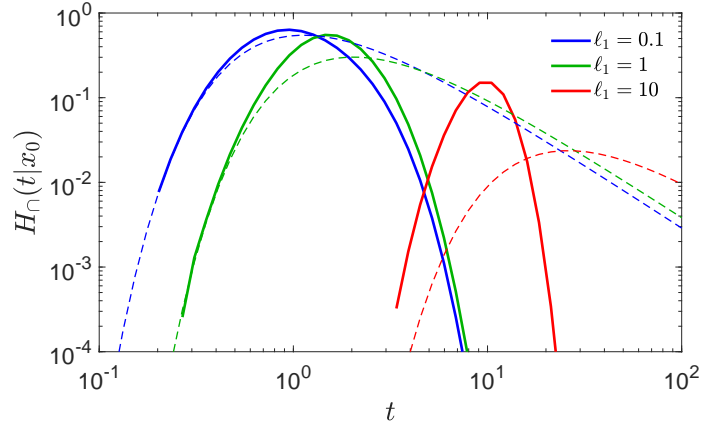} % Hcap.eps}
\end{center}
\caption{
The probability density $H_{\cap}(t|x_0)$ of the first-crossing time
$\tau_{\cap}$, with $b = 1$, $D = 1$, $x_0 = 0.5$, $\ell_2 = 1$, and
three values of $\ell_1$ as indicated in the plot.  Solid lines show
the numerical inversion of $\tilde{H}_\cap(p|x_0)$ via the Talbot
algorithm, whereas dashed lines indicate the short-time asymptotic
relation (\ref{eq:Hcap_short}).  Some points are missing at short
times due to instabilities of the numerical inversion of the Laplace
transform.}
\label{fig:Ht_cap}
% A_localtime3_FH_t_fig3(rhot1,rhot2,rhot3);
\end{figure}

Figure \ref{fig:Ht_cap} illustrates the behavior of the probability
density $H_{\cap}(t|x_0)$.  As previously for $H_{\cup}(t|x_0)$, the
short-time asymptotic relation (\ref{eq:Hcap_short}) is accurate for
small and moderate $\ell$, while its range of applicability is limited
for large $\ell$.  Expectedly, all curves are shifted to longer times
as compared to Fig. \ref{fig:Ht_cup} because the condition determining
$\tau_{\cap}$ is more strict than that determining $\tau_{\cup}$.

\section{Discussion and conclusion}
\label{sec:conclusion}

In this paper, we extended the approach relying on the concept of the
boundary local time as a proxy for the number of encounters with the
boundary that was recently developed to describe diffusion-mediated
surface phenomena \cite{Grebenkov20}.  Our extension allows one to
partition the boundary into regions with distinct reactivities and to
characterize encounters with each region.  For this purpose, we
introduced the full propagator $P(\x,\ell_1,\ldots,\ell_m,t|\x_0)$ as
the joint probability density of the position of the particle and of
its multiple boundary local times on each boundary region.  This
propagator was then related via Eq. (\ref{eq:GP_general}) to the
conventional propagator $G_{q_1,\ldots,q_m}(\x,t|\x_0)$ satisfying
Robin boundary conditions with parameters $q_1,\ldots,q_m$ on boundary
regions $\Gamma_i$.  The explicit implementation of the surface
reactivities via the factors $e^{-q_1\ell_1} \ldots e^{-q_m\ell_m}$ in
the expression (\ref{eq:GP_general}) opens a way to investigate
various surface reaction mechanisms such as, e.g., catalysts' fooling
or membrane degradation \cite{Bartholomew01,Filoche08}.  In fact, the
parameters $q_i$ enter into the conventional propagator
$G_{q_1,\ldots,q_m}(\x,t|\x_0)$ via the Robin boundary condition
(\ref{eq:Robin}) that corresponds to the Poissonian type of surface
reaction: at each encounter with $\Gamma_i$, the probability of the
reaction event is the same.  The factor $e^{-q_i\ell_i}$ is thus the
probability of no surface reaction on $\Gamma_i$, i.e., the
probability $\P\{\hat{\ell}_i > \ell_t^i\}$ that the boundary local
time $\ell_t^i$ does not exceed its random threshold $\hat{\ell}_i$
obeying the exponential distribution with the mean $1/q_i$.  However,
one can go beyond this conventional choice and consider a variety of
surface reaction mechanisms characterized by any desired distribution
of the threshold $\hat{\ell}_i$: $\P\{\hat{\ell}_i > \ell_i\} =
\Psi_i(\ell_i)$ (see \cite{Grebenkov20} for details).  The generalized
propagator describing the likelihood of finding the particle in $\x$
survived against such surface reactions will then be
\begin{equation}
G_{\rm gen}(\x,t|\x_0) = \int\limits_0^\infty d\ell_1 \, \Psi_1(\ell_1) \ldots \int\limits_0^\infty d\ell_m \, \Psi_m(\ell_m) \,
P(\x,\ell_1,\ldots,\ell_m,t|\x_0) .
\end{equation}
In this way, we extend the approach developed in \cite{Grebenkov20} in
order to implement various surface reaction mechanisms individually
for each region $\Gamma_i$ of the boundary.  Several models of random
thresholds and their consequences on the distribution of the reaction
time were discussed in \cite{Grebenkov20}.  An interesting perspective
consists in studying these models in the current setting with multiple
boundary local times (and thus multiple thresholds $\hat{\ell}_i$).
The exact formula (\ref{eq:Pfull}) for the full propagator on the
interval and its extensions to a circular annulus and a spherical
shell will be particularly helpful.  

Another interesting extension consists in studying the limit $m\to
\infty$ of finer and finer partitions of the boundary $\pa$.  As a
sequence of piecewise constant functions can approximate a given
function $q_{\s}$ characterizing the reactivity of the boundary, one
can access the general case of a space-dependent reactivity, in which
the propagator $G_{q_{\s}}(\x,t|\x_0)$ satisfies the Robin boundary
condition:
\begin{equation}
\bigl(\partial_n G_{q_{\s}}(\x,t|\x_0)\bigr)_{\x = \s} + q_{\s} \, G_{q_{\s}}(\s,t|\x_0) = 0  \qquad (\s \in \pa).
\end{equation}
Indeed, Eq. (\ref{eq:GP_general}) can formally be written as a sort of
Feynman's path integral (here, we do not provide any rigorous
statements but just sketch the main ideas):
\begin{subequations}
\begin{eqnarray}
G_{q_{\s}}(\x,t|\x_0) &=& \int {\mathcal D} \ell_{\s} \, \exp\biggl(-\int\nolimits_{\pa} d\s \, q_{\s} \, \ell_{\s}\biggr) P(\x,\ell_{\s},t|\x_0) \\
&=& \E_{\x_0} \left\{ \exp\biggl(-\int\nolimits_{\pa} d\s \, q_{\s} \, \ell_t^{\s}\biggr) \, \delta(\X_t - \x) \right\} ,
\end{eqnarray}
\end{subequations}
where $\ell_t^{\s}$ is the boundary local time in an infinitesimal
vicinity of the boundary point $\s$.  As $\ell_t^{\s}$ increases only
when the particle hits a vicinity of the point $\s$, the integral over
$\s$ can be re-arranged as
\begin{equation}
\int\nolimits_{\pa} d\s \, q_{\s} \, \ell_t^{\s} = \int\limits_0^t q_{\X_{t'}}\, d\ell_{t'} \,,
\end{equation}
where $d\ell_{t'}$ denotes increments of the total boundary local time
$\ell_t$ on the whole boundary $\pa$.  Using this relation, one gets
a probabilistic representation
\begin{equation}
G_{q_{\s}}(\x,t|\x_0) = \E_{\x_0} \left\{ \exp\biggl(- \int\limits_0^t q_{\X_{t'}}\, d\ell_{t'} \biggr) \, \delta(\X_t - \x) \right\} ,
\end{equation}
which is more conventional for the mathematical literature on
stochastic processes \cite{Papanicolaou90,Bass08}.  On the other hand,
a spectral expansion of the propagator $G_{q_{\s}}(\x,t|\x_0)$ in
terms of the eigenfunctions of the operator $\M_p + q_{\s}$ was
derived in \cite{Grebenkov19}.  Further mathematical analysis of this
intricate relation presents an interesting perspective for future
research.

We also discussed a variety of the first-passage times associated to
this problem.  After identifying the conventional cases of perfectly
and partially reactive targets, we introduced a new class of
first-passage times characterizing the moment of the first crossing of
prescribed thresholds by two boundary local times.  We derived the
exact formulas for the Laplace-transformed probability densities of
such first-crossing times $\tau_{\cup}$ and $\tau_{\cap}$.  We also
analyzed their short-time asymptotic behavior and obtained the mean
values of these random variables.  In turn, getting the long-time
asymptotic behavior, which is usually much simpler for first-passage
times, remains an open problem (see also discussion in
\ref{sec:A_FPT}).  Further progress in this direction may
potentially be achieved with the help of the Donsker-Varadhan large
deviation theory \cite{Donsker75,Angeletti16}.
The obtained probability densities of the first-passage times could
then be used for implementing new surface reaction mechanisms via
stopping conditions.
Note that we focused on first-passage times related to the joint
probability density $P(\circ, \ell_1,\ell_2,t|x_0)$ of two boundary
local times.  Another perspective consists in extending the obtained
results by using the full propagator $P(x, \ell_1,\ell_2,t|x_0)$ and
thus conditioning on the arrival point.

While most explicit results were presented for the interval, an
extension to a circular annulus and a spherical shell is
straightforward.  All three domains are often used as models in
various physical, chemical and biological applications.  For instance,
diffusion in an interval can model diffusion-influenced reactions in
layered structures (such as a slab); diffusion in a circular annulus
can be relevant for cylinder-shaped confinements (e.g., the interior
space of a bacterium which contains nucleotides in the middle and is
surrounded by an outer membrane); similarly, diffusion in a spherical
shell can model diffusive processes inside the cytosol surrounded by
the cellular and nuclear membranes.  Apart from these basic models and
related applications, the analytical results of the paper shed a light
on the elaborate statistics of two boundary local times.  In
particular, the intrinsic correlations between these two processes
illustrate the difficulties in getting more explicit results for
general domains.  In this perspective, the present work makes the
first steps on the way toward the full description of boundary
encounters and related surface reactions.

\section*{Acknowledgments}
The author is grateful to G. Oshanin for fruitful discussions of the
inverse Laplace transforms.  A partial financial support from the
Alexander von Humboldt Foundation through a Bessel Research Award is
acknowledged.

\appendix
\section{Dirichlet-to-Neumann operator for an interval}
\label{sec:DN}

The Dirichlet-to-Neumann operator and its spectral properties were
employed to describe diffusion-mediated surface phenomena in
\cite{Grebenkov20} (see also \cite{Grebenkov19}).  For a domain
$\Omega\subset \R^d$ with a smooth boundary $\pa$, the
Dirichlet-to-Neumann operator $\M_p$ associates to each (appropriate)
function $f$ on the boundary $\pa$ another function $g$ on that
boundary such that $\M_p f = g = (\partial_n w)|_{\pa}$, where $w(\x)$
is the solution of the modified Helmholtz equation $(p-D\Delta)w(\x) =
0$ in $\Omega$ with Dirichlet boundary condition $w|_{\pa} = f$.  In
other words, the operator $\M_p$ maps Dirichlet boundary condition
$w|_{\pa} = f$ to Neumann boundary condition $(\partial_n w)|_{\pa} =
g = \M_p f$ for the same solution $w(\x)$ (see
\cite{Grebenkov19,Grebenkov20} for further discussion and references).

A general solution of the modified Helmholtz equation on an interval
$(0,b)$ can be written as
\begin{equation}  \label{eq:wx}
w(x) = c_1 \, \frac{\sinh (\alpha(b-x))}{\sinh (\alpha b)} + c_2 \, \frac{\sinh(\alpha x)}{\sinh (\alpha b)} \,,
\end{equation}
with unknown coefficients $c_1$ and $c_2$.  As any ``function'' on the
boundary of the interval can be represented by a two-dimensional
vector $(f_1,f_2)^\dagger$ (with coefficients $f_1$ and $f_2$), the
Dirichlet-to-Neumann operator acts here as a $2\times 2$ matrix
\begin{equation}
\M_p f = \left(\begin{array}{cc} \alpha \, \ctanh(\alpha b) &  - \alpha/\sinh(\alpha b) \\
- \alpha/\sinh(\alpha b) & \alpha \, \ctanh(\alpha b) \end{array} \right)
\left(\begin{array}{c} f_1 \\ f_2 \end{array} \right) .
\end{equation}
One can recognize the parameters $C$ and $E/2$ from Eqs. (\ref{eq:C},
\ref{eq:E}) as the diagonal and non-diagonal elements of this matrix.
The eigenvalues and eigenvectors of this matrix are:
\begin{subequations}
\begin{align}
\mu_1 & = C - E/2 = \alpha \, \tanh(\alpha b/2), \quad v_1 = \frac{1}{\sqrt{2}} (1,~ 1)^\dagger  , \\
\mu_2 & = C + E/2 =  \alpha \, \ctanh(\alpha b/2) , \quad v_2 = \frac{1}{\sqrt{2}} (1, ~ -1)^\dagger  .
\end{align}
\end{subequations}
Using Eqs. (\ref{eq:jinfty}), one finds
\begin{align}
V_1^{(p)}(x_0) & = \frac{\sinh(\alpha(b-x_0)) + \sinh(\alpha x_0)}{\sqrt{2} \, \sinh(\alpha b)} \,, \\
V_2^{(p)}(x_0) & = \frac{\sinh(\alpha(b-x_0)) - \sinh(\alpha x_0)}{\sqrt{2} \, \sinh(\alpha b)} \,,
\end{align}
where $V_n^{(p)}(x_0)$ were defined in \cite{Grebenkov20} as
projections of $\tilde{j}_{\infty,\infty}(s,p|x_0)$ onto the
eigenfunctions of the Dirichlet-to-Neumann operator.

Using these expressions, we can compute the full propagator
$\tilde{P}_{\rm tot}(x,\ell,p|x_0)$ in the case of equal reactivities
($q_1 = q_2$), which characterizes the total boundary local time
$\ell_t = \ell_t^1 + \ell_t^2$
\cite{Grebenkov20}:
\begin{align} \nonumber
& D\tilde{P}_{\rm tot}(x,\ell,p|x_0)  = D\tilde{G}_{\infty,\infty}(x,p|x_0) \delta(\ell) + \sum\limits_n V_n^{(p)}(x_0) V_n^{(p)}(x) 
e^{-\mu_n^{(p)} \ell} \\  \nonumber
& = D\tilde{G}_{\infty,\infty}(x,p|x_0) \delta(\ell) \\  \nonumber
& + e^{-C\ell} \biggl(\frac{\sinh(\alpha(b-x_0)) \sinh(\alpha (b-x)) +
\sinh(\alpha x_0) \sinh(\alpha x)}{\sinh^2(\alpha b)} \, \cosh(E\ell/2) \\  \label{eq:Ptot_int}
& - \frac{\sinh(\alpha(b-x_0)) \sinh(\alpha x) + \sinh(\alpha x_0) \sinh(\alpha (b-x))}{\sinh^2(\alpha b)}\, \sinh(E\ell/2) \biggr),
\end{align}
where $\tilde{G}_{\infty,\infty}(x,p|x_0)$ is given by
Eq. (\ref{eq:tildeG}).  The marginal probability density of $\ell_t$
in the Laplace domain reads
\begin{align} \nonumber
\tilde{P}_{\rm tot}(\circ,\ell,p|x_0) & = \tilde{S}_{\infty,\infty}(p|x_0) \delta(\ell)  \\  \label{eq:Stot_int}
& + \frac{\cosh(\alpha b)-1}{\alpha \sinh(\alpha b)} \, \frac{\sinh(\alpha x_0) + \sinh(\alpha (b-x_0))}{\sinh(\alpha b)} \, \frac{e^{-(C-E/2)\ell}}{D} \,.
\end{align}
The cumulative probability function of $\ell_t$ is
\begin{equation}  \label{eq:Ftot}
%\fl
\tilde{F}_{\rm tot}(\ell,p|x_0) = \frac{1}{D\alpha^2} \biggl(\Theta(\ell) + \frac{\sinh(\alpha x_0) + \sinh(\alpha (b-x_0))}{\sinh(\alpha b)}
\bigl(1-\Theta(\ell) - e^{-(C-E/2)\ell}\bigr) \biggr),
\end{equation}
where the derivative of the Heaviside function $\Theta(\ell)$ yields
$\delta(\ell)$ in the above probability density.

\vskip 2mm

In addition to the above Dirichlet-to-Neumann operator, one can
consider other versions of this operator, which can give complementary
insights on this problem.  The first one consists in restricting the
operator to one endpoint, e.g., on $\Gamma_2 = \{b\}$.  In other
words, the modified operator acts on functions defined only on
$\Gamma_2$ (here, as the boundary $\Gamma_2$ consists of one point,
this ``functional'' space is one-dimensional), while the solution is
fixed to $0$ at the other endpoint.  This is equivalent to fixing $c_1
\equiv 0$ in Eq. (\ref{eq:wx}), and the action of the modified
Dirichlet-to-Neumann operator reads 
\begin{equation}
\M_p^D f = \left. \biggl(\partial_n \frac{\sinh(\alpha x)}{\sinh(\alpha b)} f \biggr)\right|_{x=b} = \alpha \, \ctanh(\alpha b) \, f ,
\end{equation}
where $\alpha \, \ctanh(\alpha b)$ can be interpreted as the
eigenvalue of this operator (corresponding to the eigenfunction $v =
1$).  As the space of ``functions'' is one-dimensional (i.e., the
``function'' $f$ is just a scalar), this is the only eigenvalue of the
operator.

The second modification consists in imposing Neumann boundary
condition on one endpoint, e.g., on $\Gamma_1 = \{0\}$.  A general
solution of the modified Helmholtz equation with Neumann condition at
$x = 0$ and Dirichlet condition at $x = b$ reads
\begin{equation}
w(x) = c \, \frac{\cosh(\alpha x)}{\cosh (\alpha b)} \,,
\end{equation}
and the action of the modified Dirichlet-to-Neumann operator on a
``function'' $f$ on $\Gamma_2$ is
\begin{equation}
\M_p^N f = \left. \biggl(\partial_n \frac{\cosh(\alpha x)}{\cosh(\alpha b)} f \biggr)\right|_{x=b} = \alpha  \tanh(\alpha b) \, f .
\end{equation}
Here, $\alpha  \tanh(\alpha b)$ is the eigenvalue of this operator
corresponding to the eigenfunction $v = 1$.

\section{Some properties of functions $Q_1$ and $Q_2$}
\label{sec:Q1Q2}

The functions $Q_1(z;a)$ and $Q_2(z_1,z_2;a)$ can be computed
numerically from their definition in Eqs. (\ref{eq:Q1Q2}).  In this
Appendix, we provide some additional representations and asymptotic
properties.

Using the representation:
\begin{equation}
I_0(z) = \frac{1}{\pi} \int\limits_0^\pi d\theta \, \exp(x\cos\theta) \,,
\end{equation}
one can write
\begin{align}
Q_1(z;a) & = e^{-a^2} \int\limits_0^z dx \, e^{-x} \, I_0(2a\sqrt{x}) = e^{-a^2} \int\limits_0^z dx \, e^{-x} 
\sum\limits_{n=0}^\infty \frac{(2a)^{2n}}{(2n)!} c_n \, x^n ,
\end{align}
where
\begin{equation}
c_n = \frac{1}{\pi} \int\limits_0^\pi d\theta \, [\cos(\theta)]^{2n} = \frac{(2n-1)!!}{2^n \, n!} \,.
\end{equation}
We get then
\begin{align}  \label{eq:Q1_sum}
Q_1(z;a) & = e^{-a^2}\sum\limits_{n=0}^\infty \frac{(2a)^{2n}}{(2n)!} c_n \, \biggl( n! \, e^{-z} \sum\limits_{k=0}^n \frac{z^k}{k!}\biggr)
= e^{-z-a^2} \sum\limits_{n=0}^\infty \frac{a^{2n}}{n!} \sum\limits_{k=0}^n \frac{z^k}{k!} \,.
\end{align}
Note also that the finite sum over $k$ in Eq. (\ref{eq:Q1_sum}) can be
written in terms of the upper incomplete Gamma function so that
\begin{align}  \label{eq:Q1_A}
Q_1(z;a) = e^{-a^2} \sum\limits_{n=0}^\infty \frac{a^{2n}}{n!}\, \frac{\Gamma(n+1,z)}{n!} \,.
\end{align}

Similarly, the double integral reads
\begin{align}  \nonumber
Q_2(z_1,z_2;a) & = (1-a^2) \int\limits_0^{z_1} dx_1 \int\limits_0^{z_2} dx_2 \, e^{-x_1-x_2} I_0(2a\sqrt{x_1x_2}) \\  \label{eq:Q2_series}
& = (1-a^2) e^{-z_1-z_2}  \sum\limits_{n=0}^\infty a^{2n} \biggl(\sum\limits_{k=0}^n \frac{z_1^k}{k!}\biggr)
\biggl(\sum\limits_{k=0}^n \frac{z_2^k}{k!}\biggr).
\end{align}

For large $z$, it is convenient to write
\begin{equation}
Q_1(z;a) = 1 - e^{-a^2} \int\limits_z^\infty dz \, e^{-x} \, I_0(2a\sqrt{x}) .
\end{equation}
If in addition $a \ll 1/z$, then one can expand $I_0(z)$ in a Taylor
series to get
\begin{equation}
%\fl \qquad 
Q_1(z;a) \simeq 1 - e^{-a^2 - z} \bigl(1 + (1+z)a^2 + \ldots\bigr) \simeq 1 - e^{-z}(1 + z a^2 + O(a^4)).
\end{equation}
In the limit $p\to \infty$, one gets then
\begin{align}
Q_1\bigl(C\ell_2; \sqrt{C\ell_1} \, \sech(\alpha b)\bigr) & \simeq 1 - e^{-\alpha \ell_2} + O(e^{-2\alpha b}) ,\\
Q_1\bigl(C\ell_1; \sqrt{C\ell_2} \, \sech(\alpha b)\bigr) & \simeq 1 - e^{-\alpha \ell_1} + O(e^{-2\alpha b}) .
\end{align}

Noting that
\begin{equation}
Q_2(z_1,\infty;a) = 1 - e^{-(1-a^2)z_1} , \qquad   Q_2(\infty,z_2;a) = 1 - e^{-(1-a^2)z_2} ,
\end{equation}
one gets for very small $a$:
\begin{align*}  \nonumber
& Q_2(z_1,z_2;a) = 1 - e^{-(1-a^2)z_1} - e^{-(1-a^2)z_2} 
+ (1-a^2)\int\limits_{z_1}^\infty dx_1 \int\limits_{z_2}^\infty dx_2 \, e^{-x_1-x_2} I_0(2a\sqrt{x_1x_2}) \\  \nonumber
& \approx 1 - e^{-(1-a^2)z_1} - e^{-(1-a^2)z_2} 
+ (1-a^2) e^{-z_1-z_2} \bigl(1 + a^2(1+z_1)(1+z_2) + O(a^4)\bigr) \\
& = (1-e^{-z_1})(1-e^{-z_2}) - \bigl(z_1 e^{-z_1} + z_2 e^{-z_2} - (z_1z_2 + z_1+z_2) e^{-z_1-z_2}\bigr) a^2 + O(a^4) .
\end{align*}
In the limit $p\to \infty$, one has 
\begin{equation}
%\fl \qquad 
Q_2\bigl(C\ell_1,C\ell_2; \sech(\alpha b)\bigr) \simeq 1 - e^{-\alpha \ell_1 } - e^{-\alpha \ell_2} 
+ e^{-\alpha(\ell_1 + \ell_2)} + O(e^{-2\alpha b})\,.
\end{equation}

\section{Two Laplace transform inversion formulas}
\label{sec:Laplace_inv}

In this Appendix, we aim at computing two classes of the inverse
Laplace transform:
\begin{subequations}
\begin{eqnarray}
U(t) &=& \L^{-1} \bigl\{ \exp(-x \, f(e^{-a \sqrt{p}}))\bigr\} ,  \\
V(t) &=& \L^{-1} \bigl\{ \exp(-x \, a \sqrt{p} \, f(e^{-a \sqrt{p}}))\bigr\} ,
\end{eqnarray}
\end{subequations}
where $a > 0$, $x > 0$, and $f(z)$ is an analytic function.

The first step consists in replacing $a \sqrt{p}$ by $p$ with the help
of the following identity 
\begin{equation}
%\fl \qquad 
\L \biggl\{  \int\limits_0^\infty d\tau \, F(t,\tau)  \, h(\tau) \biggr\}(p) 
= \int\limits_0^\infty d\tau \, e^{-a \tau \sqrt{p}} \, h(\tau) = \L\{h\}(a\sqrt{p}) = \tilde{h}(a\sqrt{p}),
\end{equation}
where $h(t)$ is a given function, and
\begin{equation}
F(t,\tau) = \frac{a \tau}{\sqrt{4\pi t^3}} \,  e^{-a^2 \tau^2/(4t)} \,.
\end{equation}
Inverting this identity, we get another identity for a given function
$\tilde{h}(p)$:
\begin{equation}
\L^{-1}\{ \tilde{h}(a \sqrt{p})\} = \int\limits_0^\infty d\tau \, F(t,\tau) \, \L^{-1}\{\tilde{h}\}(\tau) .
\end{equation}
Using this representation, we have
\begin{equation}
U(t) = \int\limits_0^\infty d\tau \, F(t,\tau) \, \L^{-1} \bigl\{ \exp(-x \, f(e^{-p}))\bigr\}(\tau) .
\end{equation}

In the second step, we expand the exponential function and use the
Taylor series
\begin{equation}
[f(z)]^n = \sum\limits_{k=0}^\infty f_{n,k} \, z^k 
\end{equation}
to write
\begin{align*}
U(t) & = \int\limits_0^\infty d\tau \, F(t,\tau) \, \L^{-1} \biggl\{ \sum\limits_{n,k} \frac{(-x)^n}{n!} \, f_{n,k} \, e^{-kp} \biggr\}(\tau) \\
& = \sum\limits_{k=0}^\infty F(t,k) \sum\limits_{n=0}^\infty \frac{(-x)^n}{n!} \, \frac{1}{k!} \, 
\biggl(\frac{\partial^k}{\partial z^k} [f(z)]^n \biggr)_{z=0}  , 
\end{align*}
where we used that the inverse Laplace transform of $e^{-kp}$ is
$\delta(\tau - k)$.  Finally, the series over $n$ yields the
exponential function, so that we conclude
\begin{equation}
%\fl \quad 
\L^{-1} \bigl\{ \exp(-x \, f(e^{-a \sqrt{p}}))\bigr\}(t)
 = \frac{a}{\sqrt{4\pi t^3}}  \sum\limits_{k=1}^\infty \frac{e^{-a^2 k^2/(4t)}}{(k-1)!} \, 
\lim\limits_{z\to 0} \biggl(\frac{\partial^k}{\partial z^k} \exp(-x f(z)) \biggr).
\end{equation}
This partly explicit expression allows one to easily compute the
short-time behavior by keeping only the first term with $k = 1$.

In the same way, we can compute the inverse Laplace transform $V(t)$:
\begin{align*} 
V(t) & = \int\limits_0^\infty d\tau \, F(t,\tau) \L^{-1} \biggl\{\sum\limits_{n,k} \frac{(-x)^n\, p^n}{n!} \, f_{n,k}  \biggr\}(\tau) \\  \nonumber
& = \sum\limits_{n,k} \frac{(-x)^n}{n!} \, f_{n,k} \int\limits_0^\infty d\tau \, F(t,\tau) \, \delta^{(n)}(\tau - k) 
= \sum\limits_{n,k} \frac{x^n}{n!} \, f_{n,k} \biggl(\frac{\partial^n}{\partial \tau^n} F(t,\tau) \biggr)_{\tau = k}  \\
& = \sum\limits_{n,k} \frac{x^n}{n!} \, \frac{1}{k!} \biggl(\frac{\partial^k}{\partial z^k} [f(z)]^n\biggr)_{z=0} 
\biggl(\frac{\partial^n}{\partial \tau^n} F(t,\tau) \biggr)_{\tau = k}  \,,
\end{align*}
where $\delta^{(n)}(z)$ is the $n$-th derivative of the Dirac
distribution.  Next, we expand the function $F(t,\tau)$ into a Taylor
series and evaluate its derivatives with respect to $\tau$:
\begin{align*}
V(t) & = \sum\limits_{n=0}^\infty \sum\limits_{k=0}^\infty \frac{x^n}{n!} \, \frac{1}{k!} \biggl(\frac{\partial^k}{\partial z^k} [f(z)]^n\biggr)_{z=0} 
B \sum\limits_{j=0}^\infty \frac{(-A)^j}{j!} \, k^{2j+1-n} \, \frac{(2j+1)!}{(2j+1-n)!} \,,
\end{align*}
where $B = a/\sqrt{4\pi t^3}$ and $A = a^2/(4t)$ (note that some terms
in this sum are strictly zero, e.g., when $2j+1 \leq n$).  Exchanging
the order of summations over $n$ and $j$, one realizes that the sum
over $n$ is the binomial expansion:
\begin{align*}
V(t) & = B \lim\limits_{z\to 0} \sum\limits_{k=0}^\infty \frac{1}{k!} \frac{\partial^k}{\partial z^k} 
\sum\limits_{j=0}^\infty \frac{(-A)^j}{j!}   \sum\limits_{n=0}^\infty [xf(z)]^n \,  k^{2j+1-n} \frac{(2j+1)!}{n! \, (2j+1-n)!} \\
& = B \lim\limits_{z\to 0} \sum\limits_{k=0}^\infty \frac{1}{k!} \frac{\partial^k}{\partial z^k} 
\sum\limits_{j=0}^\infty \frac{(-A)^j}{j!}  \bigl(k + xf(z)\bigr)^{2j+1} \\
& = B \lim\limits_{z\to 0} \sum\limits_{k=0}^\infty \frac{1}{k!} \frac{\partial^k}{\partial z^k}  (k+xf(z)) \exp\bigl(-A(k+xf(z))^2\bigr) .
\end{align*}
We conclude that
\begin{align}  \nonumber
& \L^{-1} \bigl\{ \exp(-x \, a \sqrt{p} \, f(e^{-a \sqrt{p}}))\bigr\}(t) \\  \label{eq:Vinv}
& \qquad = \frac{a}{\sqrt{4\pi t^3}} \lim\limits_{z\to 0} \sum\limits_{k=0}^\infty 
\frac{1}{k!} \frac{\partial^k}{\partial z^k} \biggl( (k+xf(z))  e^{-a^2 (k+xf(z))^2/(4t)} \biggr).
\end{align}
Keeping only the term with $k = 0$, one gets the short-time asymptotic
behavior:
\begin{equation}
V(t) \simeq  \frac{a xf(0)}{\sqrt{4\pi t^3}} \,  e^{-a^2 x^2 [f(0)]^2/(4t)} \,.
\end{equation}

In the trivial case $f(z) = 1$, Eq. (\ref{eq:Vinv}) immediately yields
the classical expression
\begin{equation}
\L^{-1} \bigl\{ \exp(-x \, a \sqrt{p} )\bigr\}(t) = \frac{a x}{\sqrt{4\pi t^3}} \, e^{-a^2 x^2/(4t)} \,.
\end{equation}

\section{The conventional propagator in two and three dimensions}
\label{sec:2d_3d}

The Laplace-transformed conventional propagator has an explicit form
in two and three dimensions due to the separation of variables.
Following \cite{Grebenkov19g}, the radial part of the propagator in
both cases reads as
\begin{equation}
%\fl \qquad 
\tilde{G}_{q_1,q_2}(r,p|r_0) = \frac{-1}{\alpha V\, W(\alpha r_0)\, \omega(r_0)} \times
\left\{ \begin{array}{l l}  v^b(r_0) \, v^a(r)  &  \quad (a \leq r \leq r_0 \leq b), \\
v^b(r) \, v^a(r_0)  &  \quad (a \leq r_0 \leq r \leq b), \end{array} \right. 
\end{equation}
where $W(z) = \K(z) \I'(z) - \I(z) \K'(z)$, $\omega(r_0)$ is the
weighting factor,
\begin{subequations}
\begin{align}
v^a(r) & = \bigl( \alpha \K'(\alpha a) - q_1 \K(\alpha a) \bigr) \I(\alpha r) - \bigl( \alpha \I'(\alpha a) - q_1 \I(\alpha a) \bigr) \K(\alpha r), \\
v^b(r) & = \bigl( \alpha \K'(\alpha b) + q_2 \K(\alpha b) \bigr) \I(\alpha r) - \bigl( \alpha \I'(\alpha b) + q_2 \I(\alpha b) \bigr) \K(\alpha r), \\
\nonumber
V & = \bigl( \alpha \K'(\alpha a) - q_1 \K(\alpha a) \bigr) \bigl( \alpha \I'(\alpha b) + q_2 \I(\alpha b) \bigr) \\  
& - \bigl( \alpha \I'(\alpha a) - q_1 \I(\alpha a) \bigr) \bigl( \alpha \K'(\alpha b) + q_2 \K(\alpha b) \bigr) ,
\end{align}
\end{subequations}
and $\I$ and $\K$ are appropriate functions. 

In two dimensions, one has
\begin{equation}
\I(z) = I_n(z), \quad \K(z) = K_n(z), \quad W(z) = 1/z, \quad \omega(r_0) = r_0, 
\end{equation}
where $I_n(z)$ and $K_n(z)$ are modified Bessel functions of the first
and second kind, respectively.  The Laplace-transformed propagator is
then
\begin{equation}  \label{eq:G_2d}
\tilde{G}_{q_1,q_2}(\x,p|\x_0) = \frac{1}{2\pi D} \sum\limits_{n=-\infty}^\infty e^{in(\phi-\phi_0)} \, \tilde{G}_{q_1,q_2}^{(n)}(r,p|r_0) ,
\end{equation}
where $\x = (r,\phi)$ and $\x_0 = (r_0,\phi_0)$ in polar coordinates,
and the superscript $(n)$ refers to the $n$-th Fourier harmonic.

In three dimensions, one has
\begin{equation}
\I(z) = i_n(z), \quad \K(z) = k_n(z), \quad W(z) = 1/z^2, \quad \omega(r_0) = r_0^2, 
\end{equation}
where $i_n(z)$ and $k_n(z)$ are modified spherical Bessel functions of
the first and second kind, respectively.  The Laplace-transformed
propagator then reads
\begin{equation}  \label{eq:G_3d}
\tilde{G}_{q_1,q_2}(\x,p|\x_0) = \frac{1}{4\pi D} \sum\limits_{n=0}^\infty (2n+1) P_n\left(\frac{(\x \cdot \x_0)}{|\x| \, |\x_0|}\right) 
\, \tilde{G}_{q_1,q_2}^{(n)}(r,p|r_0) ,
\end{equation}
where $P_n(z)$ are Legendre polynomials, $r = |\x|$, and $r_0 =
|\x_0|$.

In both cases, the dependence of the propagator on $q_1$ and $q_2$ is
identical to that in the one-dimensional case.  As a consequence, the
inversion of the double Laplace transform with respect to $q_1$ and
$q_2$ of each radial propagator can be performed explicitly, and then
the obtained contributions can be summed up according to
Eqs. (\ref{eq:G_2d}, \ref{eq:G_3d}).

\section{First-crossing time for the total boundary local time}
\label{sec:A_FPT}

In this Appendix, we study the distribution of the first-crossing time
$\tau$ of a given threshold $\ell$ by the total boundary local time
$\ell_t = \ell_t^1 + \ell_t^2$ on the interval $(0,b)$.
As discussed in Sec. \ref{sec:FPT}, the Laplace-transformed
probability density $\tilde{H}(p|x_0)$ of $\tau$ is determined by the
Laplace-transformed cumulative probability function $\tilde{F}_{\rm
tot}(\ell,p|x_0)$ given by Eq. (\ref{eq:Ftot}):
\begin{align}  \nonumber
\E_{x_0}\{ e^{-p\tau} \} & = \tilde{H}(p|x_0) = 1 - p \tilde{F}_{\rm tot}(\ell,p|x_0)  \\  \label{eq:tildeH_tot}
& = \biggl(\frac{\sinh(\alpha x_0) + \sinh(\alpha (b-x_0))}{\sinh(\alpha b)} 
\biggr) \,  e^{- \alpha \tanh(\alpha b/2) \ell} \,,
\end{align}
where we assumed $\ell > 0$ to get a simpler expression (given that
$\ell = 0$ corresponds to the well-studied case of the first-passage
time to either of endpoints).  The series expansion of this expression
for $p\to 0$ allows one to compute the moments of $\tau$:
\begin{equation}
\E_{x_0}\{ \tau^m \} = (-1)^m \lim\limits_{p\to 0} \frac{\partial^m}{\partial p^m} \tilde{H}(p|x_0) .
\end{equation}
In particular, we find the mean and the variance as
\begin{equation}
%\fl \qquad 
\E_{x_0}\{ \tau \} = \frac{x_0(b-x_0)+ \ell b}{2D} \,,  \qquad
\sigma_\tau^2 = \frac{x_0(b-x_0)(2x_0^2 - 2bx_0+b^2) + \ell b^3}{12D^2} \,.
\end{equation}
In both expressions, the first term (without $\ell$) represents the
contribution from the first-passage time to either of endpoints,
whereas the second term accounts for multiple reflections.  Indeed,
the first-crossing time $\tau$ can be split into two independent
contributions: the first-passage time to the endpoints, and the
first-crossing time starting from the endpoint.  Setting $x_0 = 0$ to
cancel the conventional contribution from the FPT, we see that both
the mean and the variance grow linearly with $\ell$.  As a
consequence, the relative standard deviation, $\sigma_\tau/\E_{x_0}\{
\tau \} = \sqrt{b/(3\ell)}$, decreases as $\ell$ grows.

As briefly mentioned in Sec. \ref{sec:time}, the standard tools for
the Laplace transform inversion (such as the residue theorem) fail
here because the exponential function in Eq. (\ref{eq:tildeH_tot})
exhibits essential singularities.  For the sake of clarify, we set
$x_0 = 0$ and consider
\begin{equation}  \label{eq:tildeH_tot0}
\tilde{H}(p|0) = e^{- \alpha \tanh(\alpha b/2) \ell} \,.
\end{equation}
Once its inverse, $H(t|0)$, is known, $H(t|x_0)$ can be obtained as a
convolution of $H(t|0)$ with the inverse of the prefactor in
parentheses in Eq. (\ref{eq:tildeH_tot}), which is well known (and can
be easily obtained via the residue theorem).

In \ref{sec:Laplace_inv}, we derive a semi-analytical formula
(\ref{eq:Vinv}) for inverting functions such as $\tilde{H}(p|0)$.
Setting $x = \ell/b$ and $a = b/\sqrt{D}$ into this formula, we get
\begin{equation}  \label{eq:Htot_exact}
H(t|0) = \frac{1}{\sqrt{4\pi D t^3}} \lim\limits_{z\to 0} \sum\limits_{k=0}^\infty 
\frac{1}{k!} \frac{\partial^k}{\partial z^k} \biggl( (bk + \ell f(z))\,  e^{-(bk + \ell f(z))^2/(4Dt)} \biggr),
\end{equation}
where $f(z) = (1-z)/(1+z)$.  The short-time behavior of this density
is obtained by keeping only the term with $k = 0$:
\begin{equation}  \label{eq:Htot_short}
H(t|0) \simeq \frac{\ell\,  e^{-\ell^2/(4Dt)} }{\sqrt{4\pi D t^3}}    \qquad (t\to 0).
\end{equation}
In contrast, getting the long-time behavior is much more difficult.
Without solving this open problem, we provide a rough approximation,
which highlights the difficulties of the long-time limit.

\subsection*{Approximate computation in the long-time limit}

\begin{figure}
\begin{center}
\includegraphics[width=100mm]{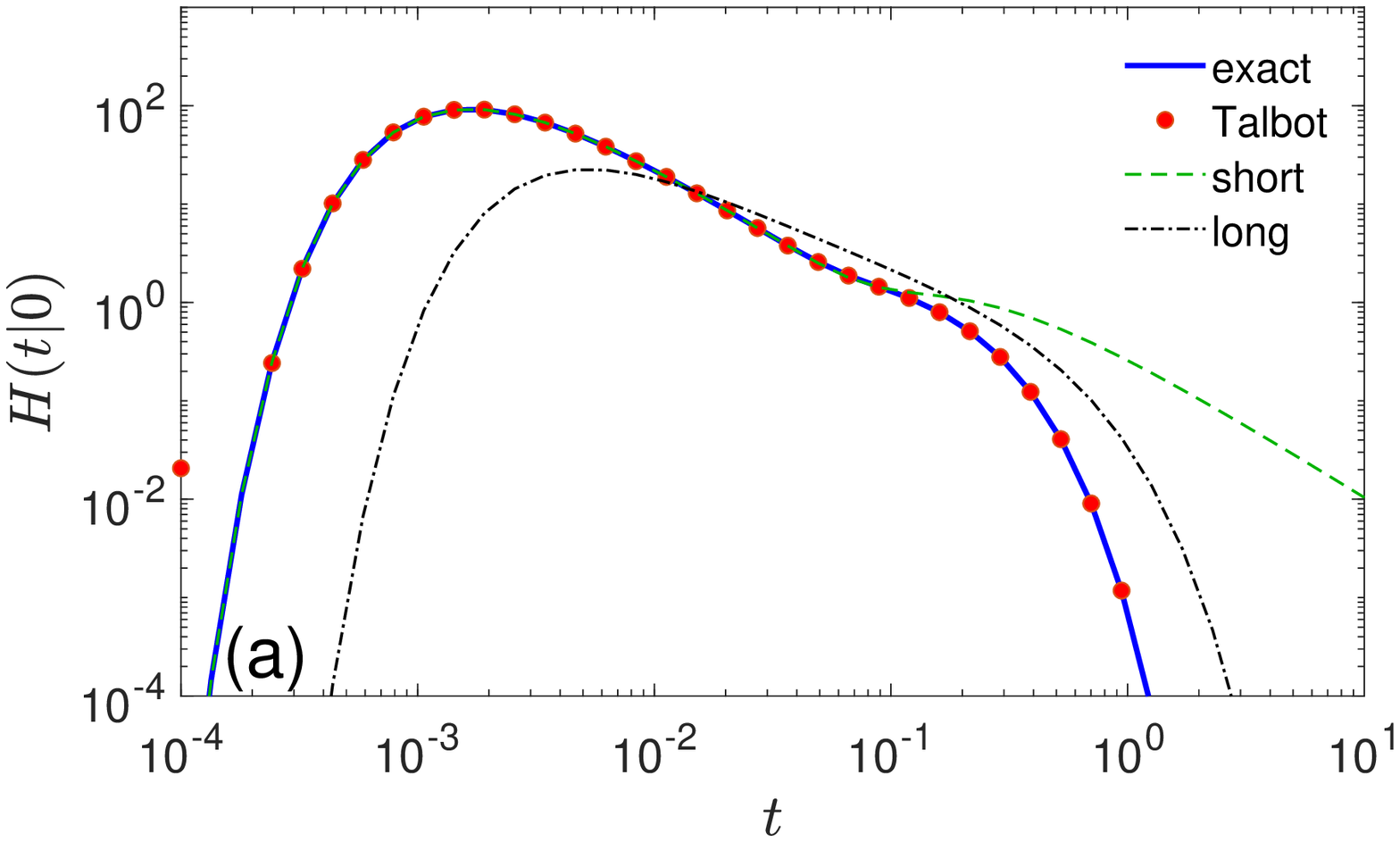} % {Htot_ell01.eps}
\includegraphics[width=100mm]{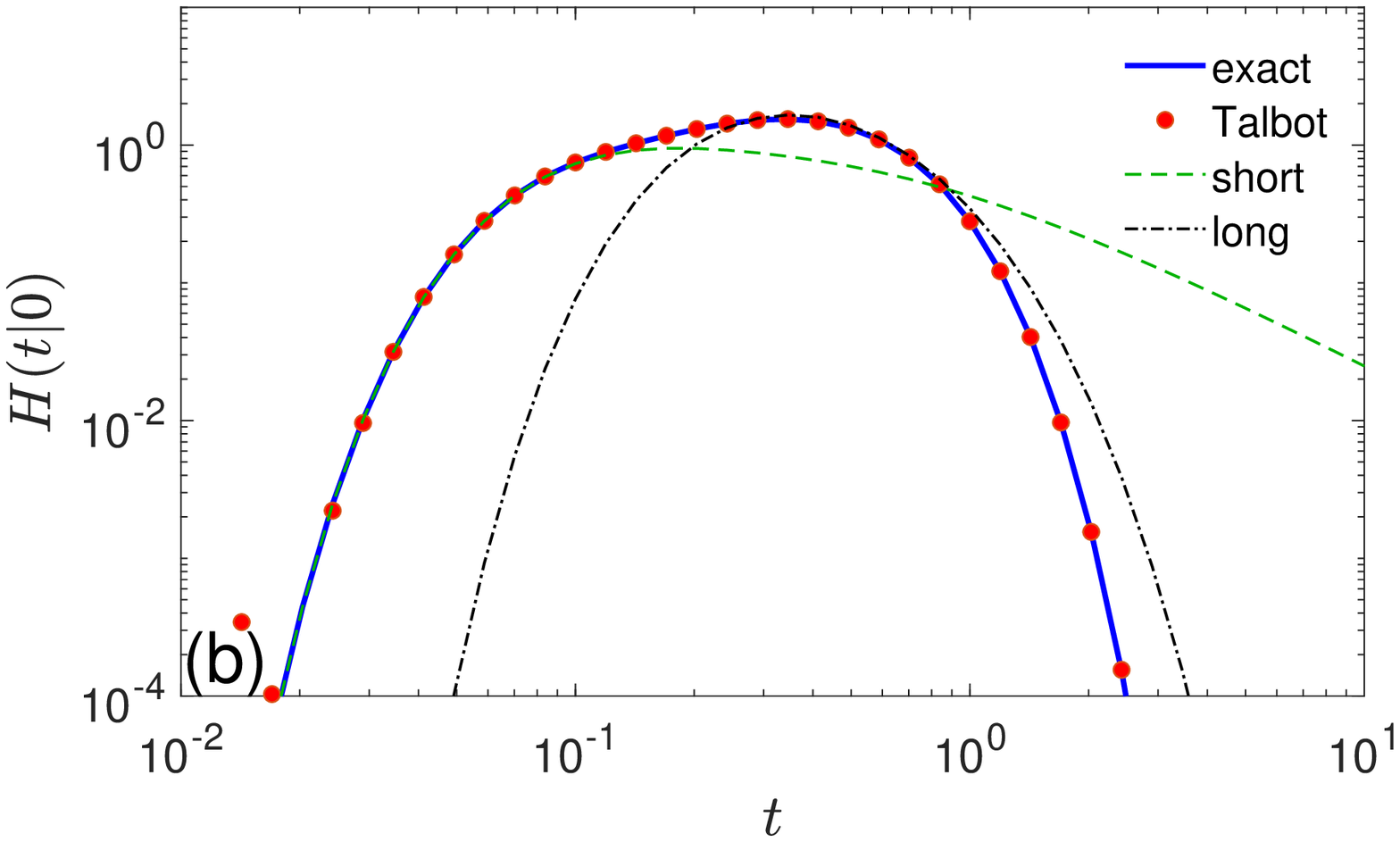} % {Htot_ell1.eps}
\includegraphics[width=100mm]{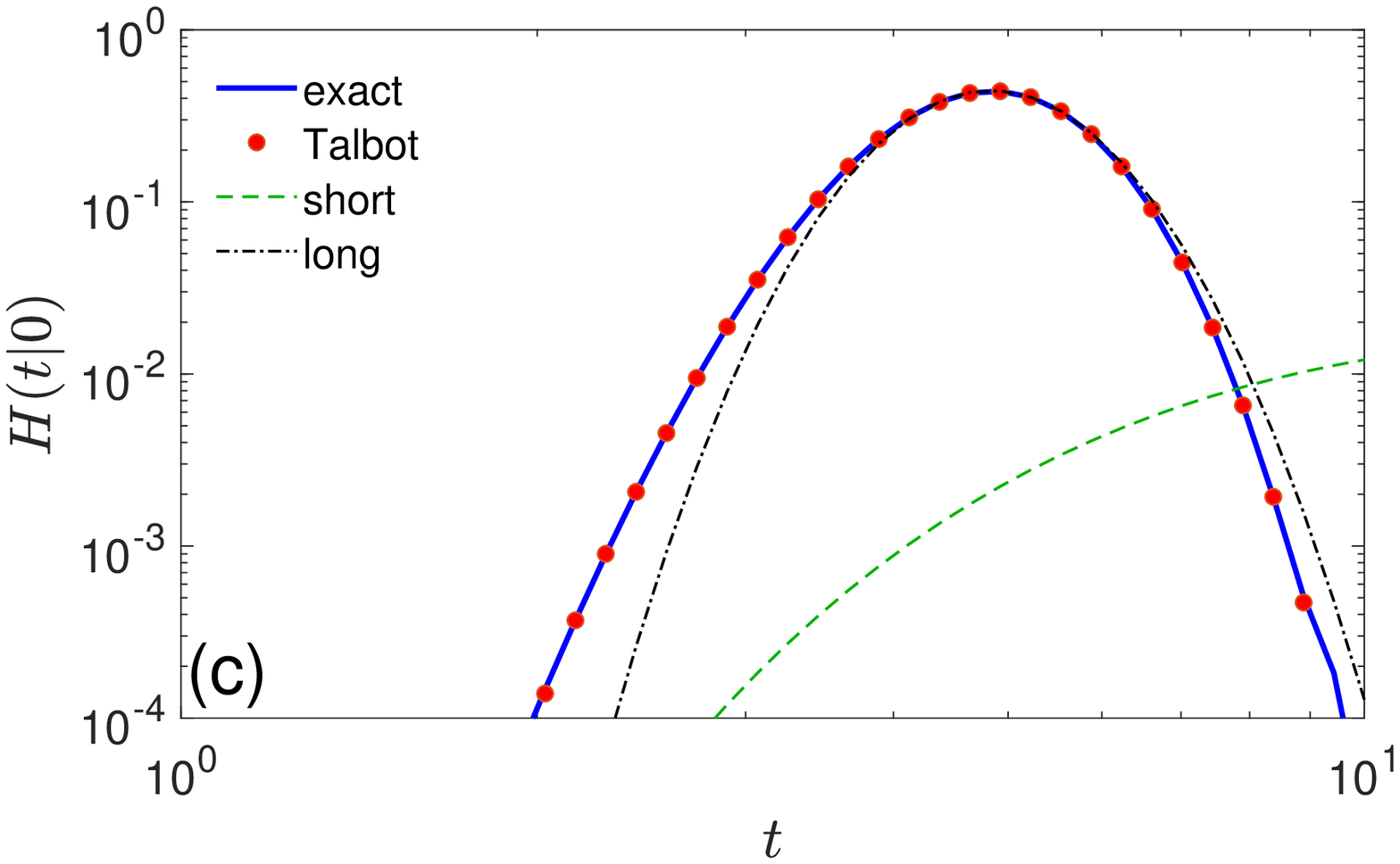} % {Htot_ell10.eps}
\end{center}
\caption{
The probability density $H(t|0)$ of the first-crossing time $\tau$ of
a threshold $\ell$ by the total boundary local time $\ell_t$ on the
interval $(0,b)$, with $b = 1$, $D = 1$, and $\ell = 0.1$ {\bf (a)},
$\ell = 1$ {\bf (b)}, and $\ell = 10$ {\bf (c)}.  Solid line presents
the exact solution (\ref{eq:Htot_exact}) truncated after $k = 20$,
filled circles show the numerical inversion by the Talbot algorithm,
dashed line indicates the short-time asymptotic relation
(\ref{eq:Htot_short}), and dash-dotted line plots the long-time
approximation (\ref{eq:Htot_long}).}
\label{fig:Htot}
% [rhot,rhotN] = A_localtime3_FHtot_t_fig2a(rhot,rhotN);   % ell = 0.1
% [rhot,rhotN] = A_localtime3_FHtot_t_fig2b(rhot,rhotN);   % ell = 1
% [rhot,rhotN] = A_localtime3_FHtot_t_fig2c(rhot,rhotN);   % ell = 10
\end{figure}

As discussed in \cite{Grebenkov07a,Grebenkov19c}, the boundary local
time in a bounded domain is close to the Gaussian distribution in the
long-time limit:
\begin{equation}
P_{\rm tot}(\circ,\ell,t|x_0) \simeq \frac{\exp(-\frac{(\ell - ct)^2}{2\beta t})}{\sqrt{2\pi \beta t}}  \,,
\end{equation}
where $c = D |\pa|/|\Omega| = 2D/b$ for an interval, and 
\begin{equation}
\beta = - \biggl(\frac{D |\pa|}{|\Omega|}\biggr)^3 \lim\limits_{p\to 0} \frac{d^2 \mu_p^{(0)}}{dp^2} = \frac{2D}{3}  \,,
\end{equation}
where we used $\mu_p^{(0)} = \alpha \, \tanh(\alpha b/2)$ for an
interval, see \ref{sec:DN}.  Note that this approximation does not
depend on the starting point $x_0$, which is irrelevant in the
long-time regime and will be omitted below.  As a consequence, we get
\begin{equation}
\P\{\tau > t\} = \P\{ \ell_t < \ell\} \simeq \frac12 \erfc\biggl(\frac{ct-\ell}{\sqrt{2\beta t}}\biggr) ,
\end{equation}
from which
\begin{align}  \label{eq:Htot_long}
H(t) & \simeq \frac{2D/b + \ell/t}{\sqrt{16\pi Dt/3}}\, \exp\biggl(-\frac{(t - b \ell/(2D))^2}{b^2 t/(3D)} \biggr)   \qquad (t\to \infty).
\end{align}

Figure \ref{fig:Htot} shows the probability density $H(t|0)$ and its
short-time and long-time approximations.  First of all, one can note
that the numerical inversion by the Talbot algorithm yields very
accurate results, with only minor deviations at short times.  As the
threshold $\ell$ increases, the distribution of the first-crossing
time is progressively shifted to longer times and becomes relatively
narrower because the relative standard deviation decreases.  For $\ell
= 0.1$ and $\ell = 1$, the short-time asymptotic formula
(\ref{eq:Htot_short}) is accurate.  In turn, for $\ell = 10$, even
though this formula is accurate at short times, the probability
density is so small due to the factor $e^{-\ell^2/(4Dt)}$ that its
range of validity is of limited interest.
In contrast, the long-time approximation (\ref{eq:Htot_long}) fails
for small and moderate $\ell$ but is getting more accurate for $\ell =
10$.

\end{document}